\documentclass[sn-basic, iicol]{sn-jnl}

\usepackage{graphicx}
\usepackage{multirow}
\usepackage{amsmath,amssymb,amsfonts}
\usepackage{amsthm}
\usepackage{mathrsfs}
\usepackage[title]{appendix}
\usepackage{xcolor}
\usepackage{textcomp}
\usepackage{manyfoot}
\usepackage{booktabs}
\usepackage{algorithm}
\usepackage{algorithmicx}
\usepackage{algpseudocode}
\usepackage{listings}
\usepackage{placeins}
\usepackage{physics}
\usepackage{multicol}
\usepackage{geometry}
\usepackage{chngcntr}
\usepackage{comment}
\usepackage[normalem]{ulem}
\usepackage{siunitx}
\usepackage{todonotes}
\begin{document}

%%%%%%%%%%%%%%%%%%%%%%%%%%%%%%%%%%%%%%%%%%%%%%%%%%%%%%%%%%%%
%%% T I T L E %%%%%%%%%%%%%%%%%%%%%%%%%%%%%%%%%%%%%%%%%%%%%%
%%%%%%%%%%%%%%%%%%%%%%%%%%%%%%%%%%%%%%%%%%%%%%%%%%%%%%%%%%%%

% title
\title[Article Title]{\textbf{A Hybrid Quantum Solver for Gaussian Process Regression}}

% authors
\author*[1]{\fnm{Kerem} \sur{B\"ukr\"u}}\email{kerem.buekrue@dlr.de}
\author[2]{\fnm{Steffen} \sur{Leger}}\email{steffen.leger@dlr.de}
\author[2]{\fnm{M. Lautaro} \sur{Hickmann}}\email{lautaro.hickmann@dlr.de}
\author[2]{\fnm{Hans-Martin} \sur{Rieser}}\email{hans-martin.rieser@dlr.de}
\author[1]{\fnm{Ralf} \sur{Sturm}}\email{ralf.sturm@dlr.de}
\author[1]{\fnm{Tjark} \sur{Siefkes}}\email{tjark.siefkes@dlr.de}

% Institute of Vehicle Concepts
\affil[1]{\orgdiv{Institute of Vehicle Concepts}, \orgname{German Aerospace Center (DLR)}, \orgaddress{\street{Pfaffenwaldring 38-40}, \city{Stuttgart}, \postcode{70569}, \country{Germany}}}

% Institute for AI Safety and Security
\affil[2]{\orgdiv{Institute for AI Safety and Security}, \orgname{German Aerospace Center (DLR)}, \orgaddress{\city{Ulm and St. Augustin},  \country{Germany}}}

%%%%%%%%%%%%%%%%%%%%%%%%%%%%%%%%%%%%%%%%%%%%%%%%%%%%%%%%%%%%
%%% A B S T R A C T %%%%%%%%%%%%%%%%%%%%%%%%%%%%%%%%%%%%%%%%
%%%%%%%%%%%%%%%%%%%%%%%%%%%%%%%%%%%%%%%%%%%%%%%%%%%%%%%%%%%%
\abstract{
Gaussian processes are widely known for their ability to provide probabilistic predictions in supervised machine learning models. Their non-parametric nature and flexibility make them particularly effective for regression tasks. However, training a Gaussian process model using standard methods requires matrix inversions with a cubic time complexity, which poses significant computational challenges for inference on larger datasets. Quantum algorithms, such as the HHL algorithm, have been proposed as solutions that overcome the need for classical matrix inversions by efficiently solving linear systems of equations using quantum computers. However, to gain a computational advantage over classical algorithms, these algorithms require fault-tolerant quantum computers with a large number of qubits, which are not yet available. The Variational Quantum Linear Solver is a hybrid quantum-classical algorithm that solves linear systems of equations by optimizing the parameters of a variational quantum circuit using a classical computer. This method is especially suitable for noisy intermediate-scale quantum computers, as it does not require many qubits. It can be used to compute the posterior distribution of a Gaussian process by reformulating the matrix inversion into a set of linear systems of equations. We empirically demonstrate that using the Variational Quantum Linear Solver to perform inference for Gaussian process regression delivers regression quality comparable to that of classical methods.
}


\maketitle
%%%%%%%%%%%%%%%%%%%%%%%%%%%%%%%%%%%%%%%%%%%%%%%%%%%%%%%%%%%%
%%% I N T R O D U C T I O N %%%%%%%%%%%%%%%%%%%%%%%%%%%%%%%%
%%%%%%%%%%%%%%%%%%%%%%%%%%%%%%%%%%%%%%%%%%%%%%%%%%%%%%%%%%%%
\section{Introduction}\label{sec1}
The goal of many real-world applications is to model complex relationships between inputs and outputs. In many cases, linear regression is often used to estimate linear relationships between variables. However, as the data becomes more nonlinear, linear regression may no longer be sufficient, as the model must be capable of identifying more complex relationships. To address this problem, supervised machine learning models can be employed, as they are able to solve regression tasks using labeled data to learn relationships between inputs and outputs. 

Deep neural networks (DNNs) are widely known for their ability to model complex relations in data. Furthermore, their trainability and ability to handle large datasets make them prominent for regression tasks. Nevertheless, they are prone to overfitting, usually do not provide any measures of uncertainty, and need a substantial amount of training data. Additionally, there are applications where obtaining large amounts of data is challenging, as each sample may require hours or even days to evaluate \citep{lualdi2023multi}. Furthermore, the black-box nature of DNNs limits their interpretability, complicating the understanding of the training process.

In contrast, Gaussian processes (GPs) can make accurate predictions with limited data \citep{Rasmussen2006Gaussian}. Additionally, they use Bayesian statistics to quantify uncertainties in predictions, signifying a level of confidence in less explored regions of the parameter space. Their non-parametric nature ensures the ability to model complex black-box functions, providing lots of flexibility. GP models have a wide range of applications including robotics \citep{10.5555/3104482.3104541}, engineering \citep{LUALDI2024111325}, healthcare \citep{10.5555/3495724.3496252} and optimization \citep{NIPS2012_05311655}.  

The output of a GP is governed by the kernel function, which includes trainable hyperparameters that are adjusted during training. This kernel function defines elements of the covariance matrix, which must be inverted to train the GP. The covariance matrix is a symmetric positive definite matrix with dimensions $N \times N$. Inverting this matrix exhibits a cubic time complexity $\mathcal{O}(N^3)$ and quadratic space complexity $\mathcal{O}(N^2)$. A common approach for Bayesian inference involves using a Cholesky decomposition, which requires $N^3/6$ floating point operations (FLOPs). Solving the resulting linear systems using the Cholesky factor then requires $N^2/2$ FLOPs. Consequently, inference becomes intractable for larger datasets containing more than $10^5$ data points. Several methods exist to address these scalability issues, primarily relying on approximation techniques. Some methods use preconditioned conjugate gradients combined with GPU parallelization \citep{10.5555/3454287.3455599, wenger2022preconditioning}. Other methods, such as stochastic variational inference, exploit stochastic optimization techniques \citep{10.5555/3023638.3023667}. However, these classical methods still face limitations when dealing with extremely large-scale problems. 

In this context, quantum algorithms offer a promising alternative for overcoming these challenges. Using quantum phenomena to simulate physics was first mentioned by \cite{feynman1982simulating}. Since then, several quantum algorithms have been developed that offer advantages over their classical counterparts in specific scenarios. The Deutsch-Josza algorithm can identify whether a function is even or odd with an exponential speedup \citep{deutsch1992rapid}. Another quantum algorithm, proposed by \cite{grover1996fast}, offers a method for rapid unstructured database search, ensuring a quadratic speedup over its classical counterpart. Using the quantum Fourier transform and Hamiltonian simulation, \cite{shor1997polynomial} was able to develop a quantum algorithm that can find the prime factors of an integer in polylogarithmic time. The HHL algorithm developed by \cite{harrow2009quantum} provides an exponential speedup over standard classical methods for solving linear systems of equations. It has a time complexity of $\mathcal{O}(s^2 \, \kappa^2 \, \text{log}(N)/\epsilon)$, meaning that it is quadratic regarding the condition number $\kappa$ of the system matrix. Furthermore, it depends highly on the desired error $\epsilon$ and the sparsity $s$. This demonstrates that one must be careful talking about logarithmic time complexities since this is only valid for special cases \citep{aaronson2015read}. The HHL algorithm was successfully applied to a linear system with dimension $N=2$ on real quantum hardware \citep{barz2014two}. A notable drawback of this algorithm is that it necessitates fault-tolerant quantum computers with a substantial number of physical qubits to be applicable in practical scenarios. Since the development of such quantum computers is still an ongoing topic, fully quantum algorithms can only be used for operations involving few logical qubits. 

A class of methods for the noisy intermediate-scale quantum era, particularly in the form of variational quantum circuits, has demonstrated significant potential for achieving near-term quantum advantages \citep{preskill2018quantum}. These circuits are widely employed in the field of quantum machine learning, with one prominent example being the variational quantum eigensolver \citep{schuld2015introduction, biamonte2017quantum, peruzzo2014variational}. Another notable method is the Variational Quantum Linear Solver (VQLS), which is used to solve linear systems of equations using a hybrid quantum-classical algorithm \citep{bravo2023variational}. In the VQLS framework, trainable parameters are optimized by minimizing a specific cost function, which is evaluated using quantum circuits. The parameters of these quantum circuits are optimized using a conventional optimization algorithm on a classical computer. Recent applications of the VQLS for a support vector machine \citep{yi2023variationalquantumlinearsolver}, computational fluid dynamics \citep{10.1063/5.0201739} or discrete finite element methods \citep{e25040580} highlight the potential of this method. 

This work introduces a novel approach that utilizes the VQLS to perform Bayesian inference in a GP model by reformulating the matrix inversion as a series of linear systems of equations. The accuracy and convergence behavior will be analyzed for a specific benchmark example. This analysis will provide insights into the accuracy and computational efficiency of the proposed approach, emphasizing its potential impact on Bayesian inference in GPs. 

Given the need for scalable inference in GPs, implementing innovative methods like the VQLS could provide significant advancements over traditional approaches, particularly for large datasets. Ultimately, this research aims to deepen our understanding of how quantum algorithms can enhance Bayesian inference techniques, paving the way for more efficient and reliable supervised machine learning models.

The remainder of this document is structured as follows: In Section \ref{sec2}, the foundational concepts of GPs and Bayesian inference are introduced. Section \ref{sec3} outlines the algorithmic structure of the VQLS, detailing its derivation and operational framework. Section \ref{sec4} addresses the implementation of the VQLS into the GP model, discussing encountered difficulties, bottlenecks, and practical limitations. Finally, Section \ref{sec5} presents the results of a VQLS-enhanced Gaussian process regression (GPR) algorithm applied to a specific example, demonstrating its effectiveness and advantages. 

%%%%%%%%%%%%%%%%%%%%%%%%%%%%%%%%%%%%%%%%%%%%%%%%%%%%%%%%%%%%
%%% G A U S S I A N   P R O C E S S E S %%%%%%%%%%%%%%%%%%%%
%%%%%%%%%%%%%%%%%%%%%%%%%%%%%%%%%%%%%%%%%%%%%%%%%%%%%%%%%%%%
\section{Gaussian processes}\label{sec2}
Supervised machine learning models utilize labeled datasets to learn the relationships between input and output values. Given a training dataset $\mathcal{D}=\{\mathbf{x}_i,y_i\}^{N-1}_{i=0}$ consisting of $N$ $d$-dimensional inputs $\mathbf{x}_i \in \mathcal{X} \subseteq \mathbb{R}^d$ and corresponding labels $y_i \in \mathcal{Y} \subseteq \mathbb{R}$, GPR performs inference on a latent function $f(\mathbf{x})$ in a non-parametric manner. The labels $y_i$ are assumed to be affected by homoscedastic noise 
\begin{equation} \label{eq:yi_fxi_epsi}
   y_i = f(\mathbf{x}_i) + \epsilon_i,
\end{equation}
where the noise is assumed to be normally distributed $\epsilon_i \sim \mathcal{N}(0,\sigma_\epsilon^2)$, with a zero mean and a noise variance $\sigma_\epsilon^2$. GPs are fully defined by their mean function and covariance function, which is often referred to as the kernel function 
\begin{align}
    m(\mathbf{x})&=\mathbb{E}[f(\mathbf{x})], \label{eq:mean}\\ k(\mathbf{x},\mathbf{x}')&=\mathbb{E}[(f(\mathbf{x})-m(\mathbf{x}))(f(\mathbf{x}')-m(\mathbf{x}'))]  \label{eq:kernel}
\end{align}
motivating the notation $f(\mathbf{x}) \sim \mathcal{GP}(m(\mathbf{x}),k(\mathbf{x},\mathbf{x}'))$ to specify the GP, with $\mathbf{x}, \mathbf{x'} \in \mathcal{X}$. The operation $\mathbb{E}$ denotes the expected value of its argument. The definition of GPs indicates that the observed values $\mathbf{y} = \{y_i\}_{i=0}^{N-1}$ and the unobserved function values $\mathbf{f}_* = {\{f(\mathbf{x}_j)}\}_{j=0}^{M-1}$, with $M$ testing data points, are following a joint multivariate normal distribution. Assuming a zero prior mean and defining the covariance matrix $\mathbf{K} \in \mathbb{R}^{N \times N}$, where each component is evaluated using the covariance function $\mathbf{K}_{ij} = k(\mathbf{x}_i, \mathbf{x}_j)$, the joint distribution can be written as
\begin{equation} \label{eq:joint_distribution}
    \begin{bmatrix}
\mathbf{y} \\
\mathbf{f}_*
\end{bmatrix}
\sim \mathcal{N} \left( \begin{bmatrix}
    \mathbf{0} \\ \mathbf{0}
\end{bmatrix}, 
\begin{bmatrix}
k(\mathbf{X}, \mathbf{X}) + \sigma_\epsilon^2 \mathbf{I} & k(\mathbf{X}, \mathbf{X}_*) \\
k(\mathbf{X}_*, \mathbf{X}) & k(\mathbf{X}_*, \mathbf{X}_*)
\end{bmatrix}
\right),
\end{equation}
with the identity matrix $\mathbf{I} \in \mathbb{R}^{N \times N}$. This includes the training data points $\mathbf{X} = \{\mathbf{x}_i\}_{i=0}^{N-1}$ and testing data points $\mathbf{X}_* = \{\mathbf{x}_j\}_{j=0}^{M-1}$. The posterior distribution is obtained by conditioning the prior on the observed data $\mathbf{f}_* | \mathbf{X}, \mathbf{y}, \mathbf{X}_* \sim \mathcal{N}(\boldsymbol{\mu}_*, \boldsymbol{\Sigma}_*)$, yielding the predictive equations for the posterior mean and covariance
\begin{align} 
    \boldsymbol{\mu}_* &= \mathbf{K}_*^\top (\mathbf{K} + \sigma_\epsilon^2 \mathbf{I})^{-1} \,\mathbf{y}, \label{eq:post_mean_cov_1} \\
    \mathbf{\Sigma}_* &= \mathbf{K}_{**} - \mathbf{K}_*^\top (\mathbf{K} + \sigma_\epsilon^2 \mathbf{I})^{-1} \, \mathbf{K}_*, \label{eq:post_mean_cov_2}
\end{align}
where we use the abbreviations $\mathbf{K} :=k(\mathbf{X}, \mathbf{X})$, $\mathbf{K} _*:= k(\mathbf{X}, \mathbf{X}_*) $ and $\mathbf{K}_{**} :=k(\mathbf{X}_*, \mathbf{X}_*)$ for simplicity. We additionally introduce a notation for the noisy covariance matrix, defined as $\widehat{\mathbf{K}} := \mathbf{K} + \sigma_\epsilon^2 \mathbf{I}$. Equations \eqref{eq:post_mean_cov_1}-\eqref{eq:post_mean_cov_2} imply that a matrix inversion is required to compute the posterior distribution. The standard procedure involves a Cholesky decomposition of the covariance matrix $\widehat{\mathbf{K}} = \mathbf{L} \mathbf{L}^\top$, with the Cholesky factor $\mathbf{L} \in \mathbb{R}^{N \times N}$, which is a lower triangular matrix. Introducing the vector $\boldsymbol{\alpha} = \widehat{\mathbf{K}}^{-1} \, \mathbf{y}$, we rewrite the predictive mean as $\boldsymbol{\mu}_* = \mathbf{K}_{*}^\top \boldsymbol{\alpha}$. We compute $\boldsymbol{\alpha}$ by solving $\mathbf{L} \mathbf{L}^\top \boldsymbol{\alpha} = \mathbf{y}$ which can be done by solving two separate triangular systems of equations. We use the backslash notation $\mathbf{L} \backslash \mathbf{y}$ to define the vector $\widetilde{\mathbf{y}}$ as the solution of the linear system $\mathbf{L} \widetilde{\mathbf{y}} = \mathbf{y}$. Next, we solve the second triangular system $\boldsymbol{\alpha} = \mathbf{L}^\top \backslash \widetilde{\mathbf{y}}$. The predictive covariance in Eq. \eqref{eq:post_mean_cov_2} can be computed by using the already computed Cholesky factor $\mathbf{L}$, which can then be used to compute the term $\mathbf{v} := \mathbf{L} \backslash \mathbf{K}_*$. This term can then be used to efficiently calculate the predictive covariance $\mathbf{\Sigma}_* = \mathbf{K}_{**} - \mathbf{v}^\top \mathbf{v}$.

The time complexity of computing the posterior distribution primarily arises from the Cholesky decomposition, which requires $N^3/6$ FLOPs. Solving the triangular systems requires $N^2/2$ computations, resulting in an overall worst-case cubic time complexity $\mathcal{O}(N^3)$. On the other hand, space complexity grows quadratically with $\mathcal{O}(N^2)$, as storing the $N \times N$ covariance matrix becomes the main bottleneck. Addressing these complexities independently emphasizes the practical limitations when scaling GPR to larger datasets. Scaling GP inference to datasets with $N > 10^5$ becomes computationally prohibitive due to the cubic time complexity, which dominates the process. These scalability challenges highlight the need for more efficient approaches. 

Optimal hyperparameters for a GP model, which are associated with the chosen kernel, are determined by maximizing the log-marginal-likelihood (LML), which is defined as
\begin{equation} \label{eq:lml}
    \log p(\mathbf{y} | \mathbf{X}) = -\frac{1}{2} \mathbf{y}^\top \boldsymbol{\alpha} - \frac{1}{2}  \log {| \widehat{\mathbf{K}}|} - \frac{N}{2} \log {2 \pi} .
\end{equation}
The first term of Eq. \eqref{eq:lml} represents how well the model fits the data, analogous to the squared error between predictions and targets. The second term penalizes overly complex models by introducing a cost for high uncertainty in predictions, discouraging overfitting. Finally, the third term normalizes the likelihood, ensuring that it is properly scaled. Together, these terms balance model fit, complexity, and stability, guiding the selection of the optimal hyperparameters for the GP. It is worth mentioning that the logarithm is used for numerical stability reasons. Maximizing the LML requires computing partial derivatives with respect to the hyperparameters $\zeta_i$. The expression for the gradient is given by
\begin{align} \label{eq:lml_grad}
    \frac{\partial}{\partial \zeta_i}  \log p(\mathbf{y} | \mathbf{X}) &= \frac{1}{2} \mathbf{y}^\top \widehat{\mathbf{K}}^{-1} \frac{\partial  \widehat{\mathbf{K}}}{\partial \zeta_i} \boldsymbol{\alpha} \notag \\ &- \frac{1}{2} \text{tr} \left( \widehat{\mathbf{K}}^{-1} \frac{\partial \widehat{\mathbf{K}}}{\partial \zeta_i} \right)
\end{align}
and must be evaluated for each hyperparameter $\zeta_i$. The repetitive evaluation of the LML and its gradient is usually the most time-consuming part of GPR. 
%%%%%%%%%%%%%%%%%%%%%%%%%%%%%%%%%%%%%%%%%%%%%%%%%%%%%%%%%%%%
%%% V Q L S %%%%%%%%%%%%%%%%%%%%%%%%%%%%%%%%%%%%%%%%%%%%%%%%
%%%%%%%%%%%%%%%%%%%%%%%%%%%%%%%%%%%%%%%%%%%%%%%%%%%%%%%%%%%%
\section{Variational Quantum Linear Solver}\label{sec3}
This section summarizes the essential steps of the VQLS algorithm, as proposed by \cite{bravo2023variational}. First, the general idea is presented. Next, the definition of the cost function and its evaluation are discussed. We use the convention that general matrices and vectors for classical objects are bold, and unitary matrices and states are non-bold.
\subsection{Overview} 
The VQLS is a hybrid quantum-classical algorithm designed to solve linear system of equations. The goal is to find the solution to an arbitrary linear system of equations $\mathbf{A} \mathbf{x} = \mathbf{b}$ for a given matrix $\mathbf{A} \in \mathbb{R}^{N \times N}$, a vector $\mathbf{b} \in \mathbb{R}^{N}$ and a solution vector $\mathbf{x} \in \mathbb{R}^{N}$ using variational quantum circuits. Given the linear system size $N$, the quantum system requires $n = \lceil \log N \rceil$ qubits to represent the solution vector $\mathbf{x}$, with $n \in \mathbb{N}$. We assume $N \in \{2^k \mid k \in \mathbb{Z}^+\}$ in this work for simplicity, leading to $n = \log N$. The Quantum Linear Systems Problem (QLSP) involves preparing the state vector $\ket{x}$ such that the unnormalized state $\ket{\psi} = A \ket{x}$ is proportional to $\ket{b}$, which is the quantum state corresponding to the vector $\mathbf{b}$. An ansatz $V$ consisting of a sequence of parameterized quantum gates is applied to the zero state to prepare the solution state $\ket{x}$, so that
\begin{equation} \label{eq:x_variational}
    \ket{x(\boldsymbol{\theta})} = V(\boldsymbol{\theta}) \ket{0} .
\end{equation}
The circuit parameters $\boldsymbol{\theta} \in \Theta$, with $\Theta$ being the parameter space, are trainable parameters of the model. A classical optimizer is used to train these circuit parameters by minimizing a specific cost function, which will be introduced at a later point. After convergence, we measure the state $\ket{x(\boldsymbol{\theta})}$ to obtain the classical solution vector $\mathbf{x}$. Given that the state is normalized, we rescale it according to $\mathbf{x} \mapsto \mathbf{x} \, \| \mathbf{b}\|/\| \mathbf{A} \mathbf{x}\|$. 

The given matrix $\mathbf{A}$ has to fulfill the following requirements: non-singularity and a finite condition number $\kappa < \infty$. However, the matrix $\mathbf{A}$ cannot be directly used in the quantum algorithm. Instead, it must be decomposed into a linear combination of $L$ unitary matrices
\begin{equation} \label{eq:mat_decomp}
    \mathbf{A} = \sum_{l=0}^{L-1} c_l A_l,
\end{equation}
with unitaries $A_l$ and complex coefficients $c_l$. The unitaries $A_l$ are represented as Pauli strings, which are tensor products of Pauli matrices $\{I, X, Y, Z\}$,  such that $A_l \in \{I, X, Y, Z\}^{\otimes n}$. The maximum number of Pauli strings increases exponentially with the matrix dimensions according to $4^n$, which becomes intractable for higher dimensions. 

The vector $\mathbf{b}$ must be embedded into a corresponding normalized quantum state $\ket{b}$, which is preferably efficiently preparable using a unitary operation $U$, such that the state can be expressed as
\begin{equation} \label{eq:b_state_prep}
    \ket{b} = U \ket{0} .
\end{equation}
The state preparation is typically accomplished using amplitude or basis embedding routines, depending on the structure of $\mathbf{b}$. 

The complete VQLS algorithm is depicted in Alg. \ref{algo_vqls}. The input parameters are the system matrix $\mathbf{A}$, which must be square and padded to the next power of two, and the right-hand side vector $\mathbf{b}$, padded accordingly. The algorithm requires a classical optimizer with a specified convergence threshold $\text{tol}$ and a maximum number of iterations $N_{\text{max}}$. 

Before starting the optimization loop, the matrix $\mathbf{A}$ must be decomposed into a set of Pauli strings. Additionally, an ansatz $V$ and a state preparation unitary $U$ must be chosen. After initializing the model parameters $\boldsymbol{\theta}$ randomly, the optimization loop begins. The parameters are iteratively adjusted until the cost function reaches the desired threshold or the maximum number of iterations has elapsed. The final parameters can then be used to obtain the final solution state $\ket{x}$. The state measurement yields the solution vector $\mathbf{x}$, which is then rescaled to obtain the correct magnitude.
\begin{algorithm}[htbp]
\caption{VQLS}\label{algo_vqls}
\begin{algorithmic}[1]
\Require $\mathbf{A} \in \mathbb{R}^{N \times N}$, $\mathbf{b} \in \mathbb{R}^N$
\State Set optimization parameters: $\text{tol}$, $N_{\text{max}}$, Optimizer
\State Define ansatz circuit $V$ \Comment{Eq. \eqref{eq:x_variational}}
\State Compute matrix decomposition $A_l$, $c_l$ \Comment{Eq. \eqref{eq:mat_decomp}}
\State Define state preparation unitary $U$  \Comment{Eq. \eqref{eq:b_state_prep}}
\State Initialize parameters $\boldsymbol{\theta}$
\For{$i = 1$ to $N_{\text{max}}$}
    \State Compute local cost function $C_L $ \Comment{Eq. \eqref{eq:local_cost_function}}
    \If {$C_L < \text{tol}$} 
        \State \Return $\boldsymbol{\theta}$
    \EndIf
    \State Compute $\nabla_{\boldsymbol{\theta}} C_L$ \Comment{Gradient}
    \State $\boldsymbol{\theta} \leftarrow \text{Optimizer}(\boldsymbol{\theta}, \nabla_{\boldsymbol{\theta}} C_L)$ \Comment{Update}
\EndFor
\State $\ket{x} \leftarrow V(\boldsymbol{\theta}) \ket{0}$ \Comment{Solution state}
\State $\mathbf{x} \leftarrow \text{Measure } \ket{x}$ \Comment{Quantum to classical}
\State $\mathbf{x} \leftarrow \mathbf{x} \, \| \mathbf{b}\| / \| \mathbf{A} \mathbf{x}\|$ \Comment{Rescaling}
\State \Return $\mathbf{x}$
\end{algorithmic}
\end{algorithm}
%%%%%%%%%%%%%%%%%%%%%
%%% COST FUNCTION %%%
%%%%%%%%%%%%%%%%%%%%%
\subsection{Cost function}
To find the optimal parameters for our ansatz circuit in Eq. \eqref{eq:x_variational}, an appropriate cost function is required. In this context, the research article by \cite{bravo2023variational} presents two distinct cost functions: a global cost function and a local cost function.
%% GLOBAL COST FUNCTION
\subsubsection{Global cost function}
As the global cost function, the overlap between the (unnormalized) projector $\ketbra{\psi}{\psi}$ and the subspace orthogonal to $\ket{b}$ is used. It can be further interpreted as the expectation value of the Hamiltonian $H_G$ for the state $\ket{x}$. This Hamiltonian is defined as
\begin{equation} \label{eq:global_hamiltonian}
    H_G = A^\dagger \qty(\mathbb{I} - \ketbra{b}{b})A,
\end{equation}
with $\mathbb{I}$ being the $N$-dimensional identity operator. The overlap can be mathematically formulated as follows
\begin{equation} \label{eq:global_cost_unnormed}
   \widehat{C}_G = \text{Tr}(\ketbra{\psi}{\psi}(\mathbb{I} - \ketbra{b}{b})) = \bra{x} H_G  \ket{x} .
\end{equation}
The cost function $\widehat{C}_G$ is small if either $\ket{\psi} \propto \ket{b}$, which yields a solution to the linear system, or if the norm of $\ket{\psi}$ is small. The latter case illustrates the size dependence of the chosen cost function. Normalizing the cost function avoids the size dependence, motivating the alternative cost function
\begin{equation} \label{eq:C_global}
    C_G = \widehat{C}_G/\braket{\psi}{\psi} = 1 - \abs{\braket{b}{\Psi}}^2 = 1 - \frac{\abs{\braket{b}{\psi}}^2}{\braket{\psi}{\psi}},
\end{equation}
where we use the normalized state vector $\ket{\Psi} = \ket{\psi} / \sqrt{\braket{\psi}{\psi}}$.

Using the matrix decomposition mentioned in Eq. \eqref{eq:mat_decomp} and the variational formulation of $\ket{x}$ shown in Eq. \eqref{eq:x_variational}, the denominator of the cost function in Eq. \eqref{eq:C_global} can be formulated as follows
\begin{equation} \label{eq:psi_psi}
    \braket{\psi}{\psi} = \sum_{l=0}^{L-1} \sum_{l'=0}^{L-1} c_l c_{l'}^* \beta_{ll'},
\end{equation}
where the following inner products must be evaluated with a set of Hadamard test circuits
\begin{equation} \label{eq:beta}
    \beta_{ll'} = \expval{V^\dagger A_{l'}^\dagger A_l V}{0} .
\end{equation}
The basic concept of the Hadamard test circuit is shown in Appendix \ref{secA1}. 

Following the same procedure, the numerator can be alternatively formulated as 
\begin{equation} \label{eq:beta_psi_norm_sq}
    \abs{\braket{b}{\psi}}^2 = \sum_{l=0}^{L-1} \sum_{l'=0}^{L-1} c_l c_{l'}^* \gamma_{ll'},
\end{equation}
which involves a product of two expectation values that can be evaluated using Hadamard tests
\begin{equation} \label{eq:gamma}
    \gamma_{ll'} = \expval{U^\dagger A_l V}{0}\expval{V^\dagger A_{l'}^\dagger U}{0} .
\end{equation}
Computing the global cost function involves evaluating the quantum circuits in Eq. \eqref{eq:beta} and Eq. \eqref{eq:gamma}. This cost function was tested in \cite{bravo2023variational}, where its trainability was limited due to the occurrence of barren plateaus, especially for a larger number of qubits, which correlates with the size of the linear system \citep{mcclean2018barren}. 
%% LOCAL COST FUNCTION
\subsubsection{Local cost function}
A local cost function was introduced by \cite{bravo2023variational} to mitigate the issue of barren plateaus in higher-dimensional linear systems
\begin{equation} \label{local_cost_function_unnormed}
    \widehat{C}_L = \expval{H_L}{x} , \qquad C_L = \widehat{C}_L/\braket{\psi}{\psi},
\end{equation}
which is now defined as the expectation value of the alternative Hamiltonian
\begin{equation} \label{eq:local_hamiltonian}
    H_L = A^\dagger U \Big( \mathbb{I} - \frac{1}{n} \sum_{q=0}^{n-1} \ketbra{0_q}{0_q} \otimes \mathbb{I}_{\bar{q}}\Big) U^\dagger A .
\end{equation}
The Hamiltonian now contains a sum, where $\ket{0_q}$ represents the zero state for qubit $q$, and $\mathbb{I}_{\bar{q}}$ is the identity operator regarding every other qubit. This Hamiltonian can be reformulated using the property
\begin{equation} \label{eq:0q_0q}
 \ketbra{0_q}{0_q} = \frac{\mathbb{I}_q + Z_q}{2},
\end{equation}
which contains the Pauli-Z operator denoted as $Z_q$. The local cost function is then defined as
\begin{equation} \label{eq:local_cost_function}
    C_L = \frac{1}{2} - \frac{1}{2n} \frac{\sum_{q=0}^{n-1}\expval{U Z_q U^\dagger}{\psi}}{\braket{\psi}{\psi}} .
\end{equation}
The denominator $\braket{\psi}{\psi}$ can be computed similarly to the global case using Eq. \eqref{eq:beta}. Each term in the sum that appears in the numerator can be computed with
\begin{equation} \label{eq:psi_U_Zq_U_psi}
   \expval{U Z_q U^\dagger}{\psi} = \sum_{l=0}^{L-1} \sum_{l'=0}^{L-1} c_l c_{l'}^* \delta_{ll'}^{(q)},
\end{equation}
where the term $\delta_{ll'}^{(q)}$ denotes the value that is computed with quantum hardware. This term can be defined as
\begin{equation} \label{eq:delta}
    \delta_{ll'}^{(q)} = \expval{V^\dagger A_{l'}^\dagger U (Z_q \otimes \mathbb{I}_{\overline{q}}) U^\dagger A_l V}{0} .
\end{equation}
%%%%%%%%%%%%%%%%%%%%%%%
%%% COST EVALUATION %%%
%%%%%%%%%%%%%%%%%%%%%%%
\subsubsection{Cost evaluation}
We have presented both global and local versions of the cost function. Both can be evaluated using a set of tailored Hadamard Test circuits, as detailed in \cite{turati2024empirical}. Without using additional symmetries, the total number of evaluations for computing the term in Eq. \eqref{eq:psi_psi} amounts to $2 L^2$. By utilizing the properties $z + \overline{z} = 2 \, \Re(z)$ $\forall z \in \mathbb{C}$ and $\bra{\phi} U^\dagger \ket{\psi} = \overline{\bra{ \psi} U \ket{\phi}}$ under the assumption that $c_l \in \mathbb{R}$ for all $l=0,\dots,L-1$, we can reformulate Eq. \eqref{eq:psi_psi} to obtain
\begin{equation} \label{eq:psi_psi_new}
    \braket{\psi}{\psi} = \sum_{l=0}^{L-1} c_l^2 +\sum_{l=0}^{L-1} \sum_{l'=l+1}^{L-1} 2 c_l c_{l'} \real \big( \expval{V^\dagger A_{l}^\dagger A_{l'} V}{0} \big),
\end{equation}
as shown in \cite{turati2024empirical}. This reformulation reduces the total Hadamard tests needed to $L(L-1)/2$. 

When employing the global cost function, it is necessary to compute the term defined in Eq. \eqref{eq:beta_psi_norm_sq}. When no additional symmetries are applied, we find that a total of $2 L^2$ Hadamard tests are required for the evaluation. By splitting the sum and changing the index, the expression can be simplified to
\begin{align} \label{eq:beta_psi_new}
&\left|\braket{b}{\psi}\right|^2 = \sum_{l=0}^{L-1} c_l^2 \abs{\expval{U^\dagger A_l V}{0}}^2 \notag \\
\quad &+ \sum_{l=0}^{L-1} \sum_{l'=l+1}^{L-1} 2 c_l c_{l'} \real \left(\expval{U^\dagger A_l V}{0}\right) \real \left(\expval{U^\dagger A_{l'} V}{0} \right) \notag \\
\quad &+ \sum_{l=0}^{L-1} \sum_{l'=l+1}^{L-1} 2 c_l c_{l'} \imaginary \left(\expval{U^\dagger A_l V}{0} \right) \imaginary \left(\expval{U^\dagger A_{l'} V}{0} \right) .
\end{align}
A more detailed derivation can be found in \cite{turati2024empirical}. This approach reduces the need for evaluations from $2L^2$ to just $2L$ Hadamard tests. 

If the local cost function is employed, we need to evaluate Eq. \eqref{eq:psi_U_Zq_U_psi} which requires $n L^2$ Hadamard tests. Reformulating the corresponding expectation value defined in Eq. \eqref{eq:delta}, using the same mathematical techniques as previously applied for the global cost function, results in
\begin{align} \label{eq:local_term_new}
&\expval{V^\dagger A^\dagger U Z_q U^\dagger A V}{0} \notag\\&= \sum_{l=0}^{L-1} c_l^2 \expval{V^\dagger A_l^\dagger U Z_q U^\dagger A_l V}{0} \notag \\
& + \sum_{l=0}^{L-1} \sum_{l'=l+1}^{L-1} 2 c_l c_{l'} \real \left( \expval{V^\dagger A_l^\dagger U Z_q U^\dagger A_{l'} V}{0} \right).
\end{align}
This reformulation reduces the number of Hadamard tests to $n L (L-1)/2$. 

We can significantly decrease the required number of Hadamard tests by employing these reformulations. The number of evaluations drops from $4 L^2$ to $L(L-1)/2 + 2L$ for the global cost function. In the case of the local cost function, it reduces from $L(L-1)/2 + n L^2$ to $L (L-1)/2 + n L (L-1)/2$. In the worst-case scenario, assuming $L=N^2$ and using the relation $n=\log(N)$, the total number of evaluations scales as $(N^4 + 3 N^2)/2$ for the global case and $(N^4 - N^2) (1+\log(N))/2$ for the local case, resulting in complexities of $\mathcal{O}(N^4)$ and $\mathcal{O}(N^4 \log(N))$, respectively. 

The local cost function increases the total number of cost function evaluations in proportion to the number of qubits, highlighting the trade-off between optimizability and computational run-time.

%%%%%%%%%%%%%%%%%%%%%%%%%%%%%%%%%%%%%%%%%%%%%%%%%%%%%%%%%%%%
%%% Q U A N T U M   G A U S S I A N   P R O C E S S E S %%%%
%%%%%%%%%%%%%%%%%%%%%%%%%%%%%%%%%%%%%%%%%%%%%%%%%%%%%%%%%%%%
\section{VQLS-GP}\label{sec4}  
In this section, we combine the concepts of GPR and the VQLS and introduce a quantum-enhanced approach for GPR. Specifically, we use the VQLS to replace the Cholesky decomposition, which is commonly used to compute the predictive mean and covariance shown in Eqs. \eqref{eq:post_mean_cov_1}-\eqref{eq:post_mean_cov_2}. We refer to this VQLS-enhanced GP model as VQLS-GP. The goal is to perform GPR using the VQLS for inference to obtain the posterior distribution. The same procedure was previously applied using the HHL algorithm \citep{zhao2019quantuma}. The matrix within the VQLS-GP model is the covariance matrix with added noise $\mathbf{A} = \widehat{\mathbf{K}}$. Two possible methods for computing the posterior distribution are introduced below. 

If the system size cannot be exactly represented using $n$ qubits, the system must be transformed by adding ones to the diagonal of matrix $\mathbf{A}$ and zeros to the vector $\mathbf{b}$
\begin{equation} \label{eq:modified_A_b}
    \Tilde{\mathbf{A}} = \begin{bmatrix} \mathbf{A} &\boldsymbol{0} \\
                                \boldsymbol{0} &\mathbf{I}_r\end{bmatrix} , \quad \Tilde{\mathbf{b}} = \begin{bmatrix} \mathbf{b} \\
                                \boldsymbol{0} \end{bmatrix},
\end{equation}
where $\mathbf{I}_r$ is an identity matrix with size $r \times r$ and $r = 2^{\lceil \log N \rceil} - N$. This modification ensures that the auxiliary variables remain zero and do not influence the solution of the linear system.
\\
\\
\textbf{Option 1} The first approach involves computing the inverse of the matrix $\widehat{\mathbf{K}}$ by solving $N$ QLSPs. We use the identity $\widehat{\mathbf{K}} \widehat{\mathbf{K}}^{-1} = \mathbf{I}$ and compute the inverse column by column. This can be achieved by using unit vectors $\mathbf{e}_i \in \mathbb{R}^N$, with $(\mathbf{e}_i)_j = 1$ if $i=j$ and $(\mathbf{e}_i)_j = 0$ otherwise. These unit vectors are the columns of the identity matrix $\mathbf{I}$ and they serve as the right-hand side $\mathbf{b}$ for each $i \in \{1,2,\dots,N\}$, such that we obtain the following linear system
\begin{equation} \label{eq:opt1}
   \widehat{\mathbf{K}} \, \text{col}_i(\widehat{\mathbf{K}}^{-1}) = \mathbf{e}_i ,
\end{equation}
where $\text{col}_i (\widehat{\mathbf{K}}^{-1}) \in \mathbb{R}^{N}$ represents the $i$-th column of the inverse of the matrix $\widehat{\mathbf{K}}$. A total of $N$ linear systems must be solved to compute the inverse of matrix $\widehat{\mathbf{K}}$. The VQLS-GP algorithm for this option is presented in Alg. \ref{algo_vqls_gpr}.
\\
\\
\textbf{Option 2} Alternatively VQLS-GP can be used to compute the necessary terms in Eqs. \eqref{eq:post_mean_cov_1}-\eqref{eq:post_mean_cov_2} directly, rather than computing the inverse. To compute the posterior mean $\boldsymbol{\mu}_*$, the matrix-vector product $ \widehat{\mathbf{K}}^{-1} \mathbf {y}$ must be evaluated. We introduce the vector $\mathbf{v} \in \mathbb{R}^{N}$ as the solution to $\mathbf{v} = \widehat{\mathbf{K}} \backslash \mathbf {y}$, so that solving the linear system
\begin{equation}  \label{eq:opt2_matvec}
    \widehat{\mathbf{K}} \mathbf{v} = \mathbf{y},
\end{equation}
using the VQLS yields the required matrix-vector product. We additionally need to evaluate the matrix-matrix product $\widehat{\mathbf{K}}^{-1} \mathbf{K}_*$ to obtain the posterior covariance. We analogously introduce an auxiliary variable $\mathbf{W} \in \mathbb{R}^{N \times M}$ and define it as the solution to $\mathbf{W} = \widehat{\mathbf{K}} \backslash \mathbf{K}_*$. Now we can compute it column-wise by solving a set of linear systems
\begin{equation}  \label{eq:opt2_matmat}
     \widehat{\mathbf{K}} \text{col}_j(\mathbf{W}) = \text{col}_j (\mathbf{K}_{*}),
\end{equation}
using the VQLS. This must be solved for each column $j=1, 2, \dots, M$ of the matrix $\mathbf{K}_* \in \mathbb{R}^{N \times M}$, resulting in $M$ linear systems for the matrix-matrix multiplication. In total we have to solve $M + 1$ linear systems. 
\\ \\
It is recommended to use option 1 if $N \ll M$ and option 2 if $M \ll N$. For the examples presented in this work, we only used option 1, which involves solving $N$ linear systems to compute the inverse of the covariance matrix. This choice was made because the number of training samples is much smaller than the number of test points in our examples. However, we also presented option 2 as an alternative for situations where $M \ll N$.
\begin{algorithm}[htbp]
\caption{VQLS-GP}\label{algo_vqls_gpr}
\begin{algorithmic}[1]
\Require $\mathbf{X}$ (inputs), $\mathbf{y}$ (targets), $\mathbf{X}_*$ (test input), $k$ (covariance function), $\sigma_\epsilon^2$ (noise level)
\State $\widehat{\mathbf{K}} \leftarrow k(\mathbf{X}, \mathbf{X}) + \sigma_\epsilon^2 \mathbf{I}$
\For{$i = 1$ to $N$}
    \State $\mathbf{e}_i \leftarrow [0, \dots, 1, \dots, 0]^\top$ \Comment{1 at index $i$}
    \State $\text{col}_i(\widehat{\mathbf{K}}^{-1}) \leftarrow \text{VQLS}(\widehat{\mathbf{K}}, \mathbf{e}_i)$
\EndFor
\State $\boldsymbol{\alpha} = \widehat{\mathbf{K}}^{-1} \mathbf{y}$
\State $\mathbf{K}_{*} \leftarrow k(\mathbf{X}, \mathbf{X}_*)$
\State $\mathbf{K}_{**} \leftarrow k(\mathbf{X}_*, \mathbf{X}_*)$
\State $\boldsymbol{\mu}_* = \mathbf{K}_{*}^{\top} \boldsymbol{\alpha}$
\State $\boldsymbol{\Sigma}_* = \mathbf{K}_{**} -  \mathbf{K}_{*}^{\top}  \widehat{\mathbf{K}}^{-1} \mathbf{K}_{*}$
\State $\log p(\mathbf{y} | \mathbf{X}) = -\frac{1}{2} \mathbf{y}^\top \boldsymbol{\alpha} - \frac{1}{2}  \log {| \widehat{\mathbf{K}}|} - \frac{N}{2} \log {2 \pi}$
\State \Return $\boldsymbol{\mu}_*$ (mean), $\boldsymbol{\Sigma}_*$ (variance), $\log p(\mathbf{y} | \mathbf{X})$ (LML)
\end{algorithmic}
\end{algorithm}
\subsection{State preparation} 
The state $\ket{b}$ must be prepared using appropriate embedding routines. To ensure that the prepared state is a valid quantum state, we must first normalize the vector $\mathbf{b}$. We then prepare the state by applying a unitary matrix $U$ during quantum circuit evaluation. It is possible to use an amplitude embedding routine to embed real-valued vectors. However, if we want to solve the linear systems shown in Eq. \eqref{eq:opt1}, we can map the unit vectors $\mathbf{e}_i$ directly to the basis state
\begin{equation} \label{eq:basis_vector_embedding}
    \ket{e_i} = U \ket{0}^{\otimes n} = \ket{i} .
\end{equation}
The unitary $U$ for preparing the state $\ket{e_i}$ can be represented as a simple operation that prepares the $n$-bit binary representation of the number $i$. The time complexity of this basis embedding is $\mathcal{O}(n)$ because we only need to perform a maximum of $n$ single gate operations. 

\subsection{Ansatz} \label{sec4_ansatzes}
The choice of the ansatz for the variational quantum circuit is crucial for the convergence of the VQLS since it defines the expressivity of the Hilbert space \citep{sim2019expressibility}. We choose a hardware efficient ansatz (HEA) for the VQLS due to its trainability and reduced circuit depth \citep{kandala2017hardware, Leone2024practicalusefulness}. This ansatz initially applies a specific sequence of gates once, followed by a sequence of gates applied across $p$ layers. The number of trainable parameters for the HEA scales linearly in the number of qubits and linearly in the number of layers, giving a total of $n(p+1)$ parameters. 

The layers of a variational quantum circuit can be intuitively compared to the number of hidden layers in a DNN. More layers ensure more expressivity, while a larger parameter Hilbert space can lead to barren plateaus. This highlights that the expressivity of the ansatz does not provide a computational advantage unless an algorithm is used which can leverage that expressivity for computational gain. The VQLS was already tested with a dynamic ansatz \citep{PhysRevA.105.012423}, where the number of layers is evolved during the optimization process.

\begin{figure}[htbp]
    \centering
    \includegraphics[width=1.0\linewidth, trim=4.2cm 2.8cm 6.7cm 4.5cm, clip]{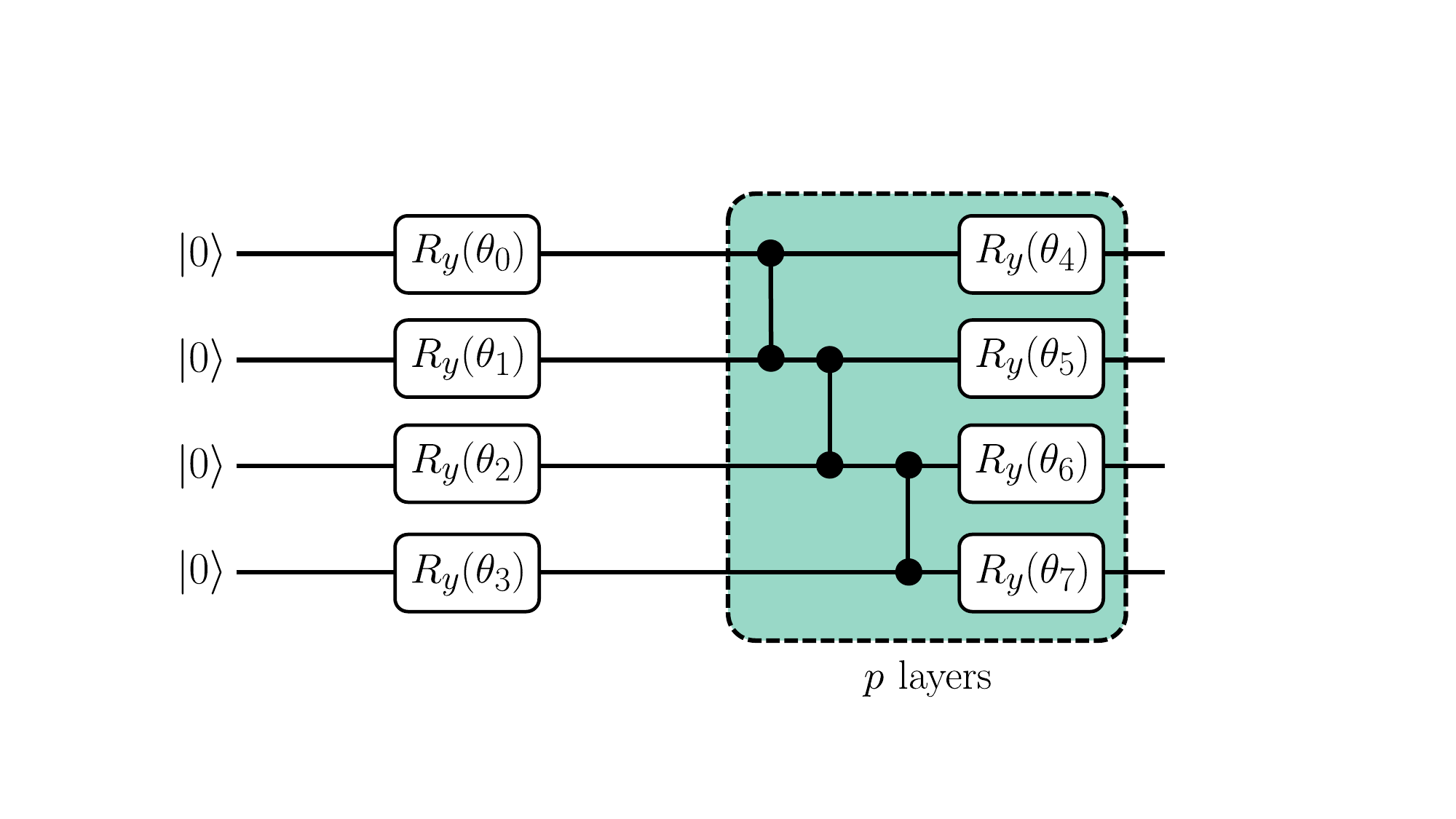}
    \caption{HEA for a four-qubit example. This ansatz consists of $n(p+1)$ trainable parameters $\boldsymbol{\theta}$ with $n$ qubits and $p$ layers}
    \label{fig:hea}
\end{figure}

We use three different ansätze for the VQLS. The first ansatz is a HEA containing rotational Y-gates and controlled Z-gates, as shown in Fig. \ref{fig:hea}. Since we only deal with real-valued linear systems of equations, we apply only Y-rotations to the solution state. At first, a rotational Y-gate is applied to each qubit. The ansatz consists of $p$ layers. In each layer, a sequence of controlled-Z (CZ) gates is applied to neighboring qubits, followed by applying a rotational Y-gate to each qubit. 

\begin{figure}[htbp] 
    \centering
    \includegraphics[width=1.0\linewidth, trim=4.2cm 2.8cm 6.7cm 4.5cm, clip]{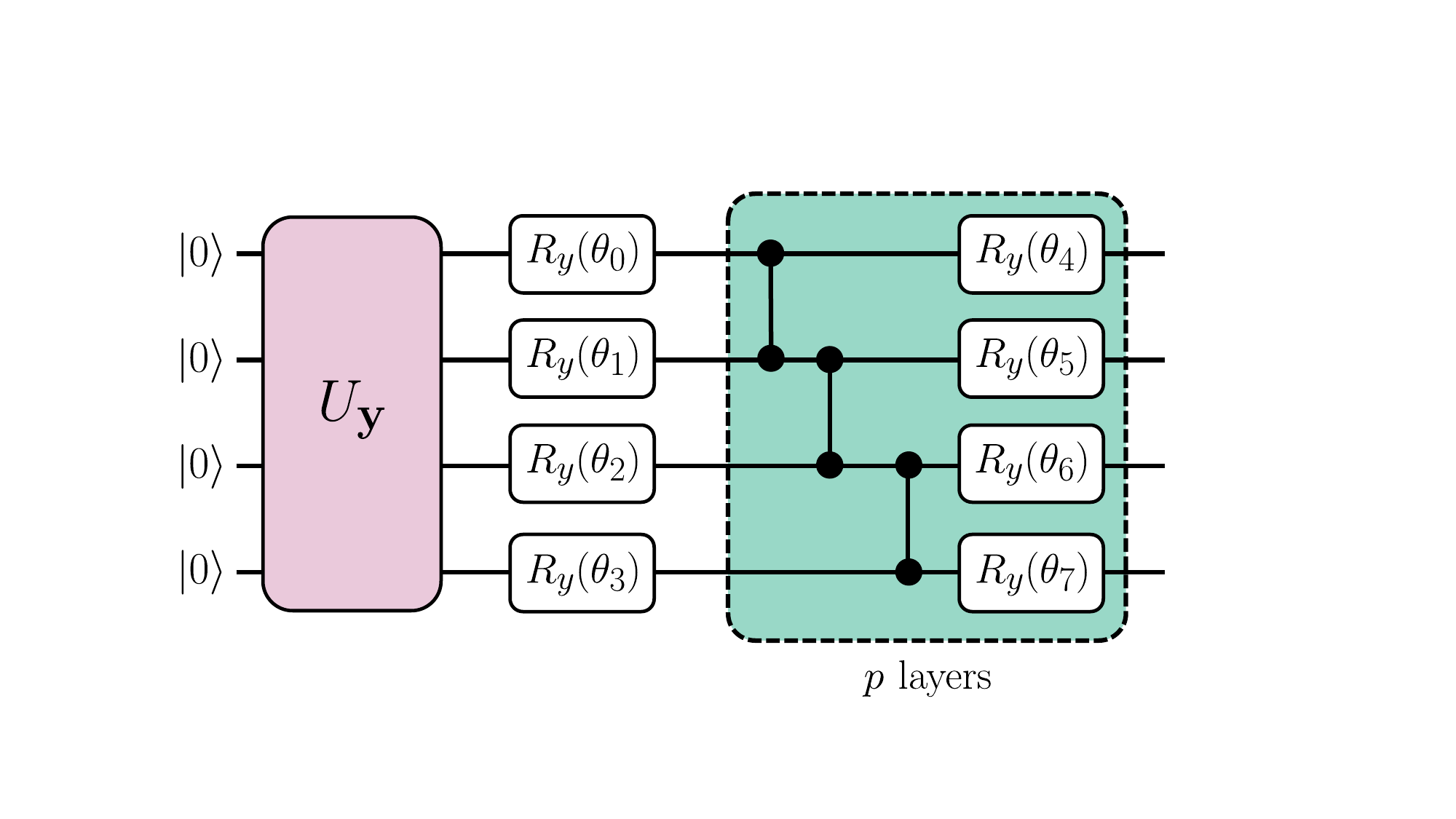}
    \caption{HEA with a single data reuploading (UHEA) for a four-qubit example. The unitary $U_{\mathbf{y}}$ embeds the training labels $\mathbf{y}$ and is applied before the HEA. The ansatz consists of $n(p+1)$ trainable parameters $\boldsymbol{\theta}$, with $n$ qubits and $p$ layers}
    \label{fig:hea_ub}
\end{figure}

The second ansatz is inspired by the HEA used in \cite{10.1063/5.0201739}. We apply a unitary $U_{\mathbf{y}}$ that embeds the training labels $\mathbf{y}$ of the linear system we are trying to solve, before the conventional HEA is applied, as presented in Fig. \ref{fig:hea_ub}. This unitary is prepared using an amplitude embedding routine introduced in \cite{mottonen2004transformation}. We refer to this ansatz as UHEA. 

\begin{figure}[htbp] 
    \centering
    \includegraphics[width=1.0\linewidth, trim=4.2cm 2.8cm 6.7cm 4.5cm, clip]{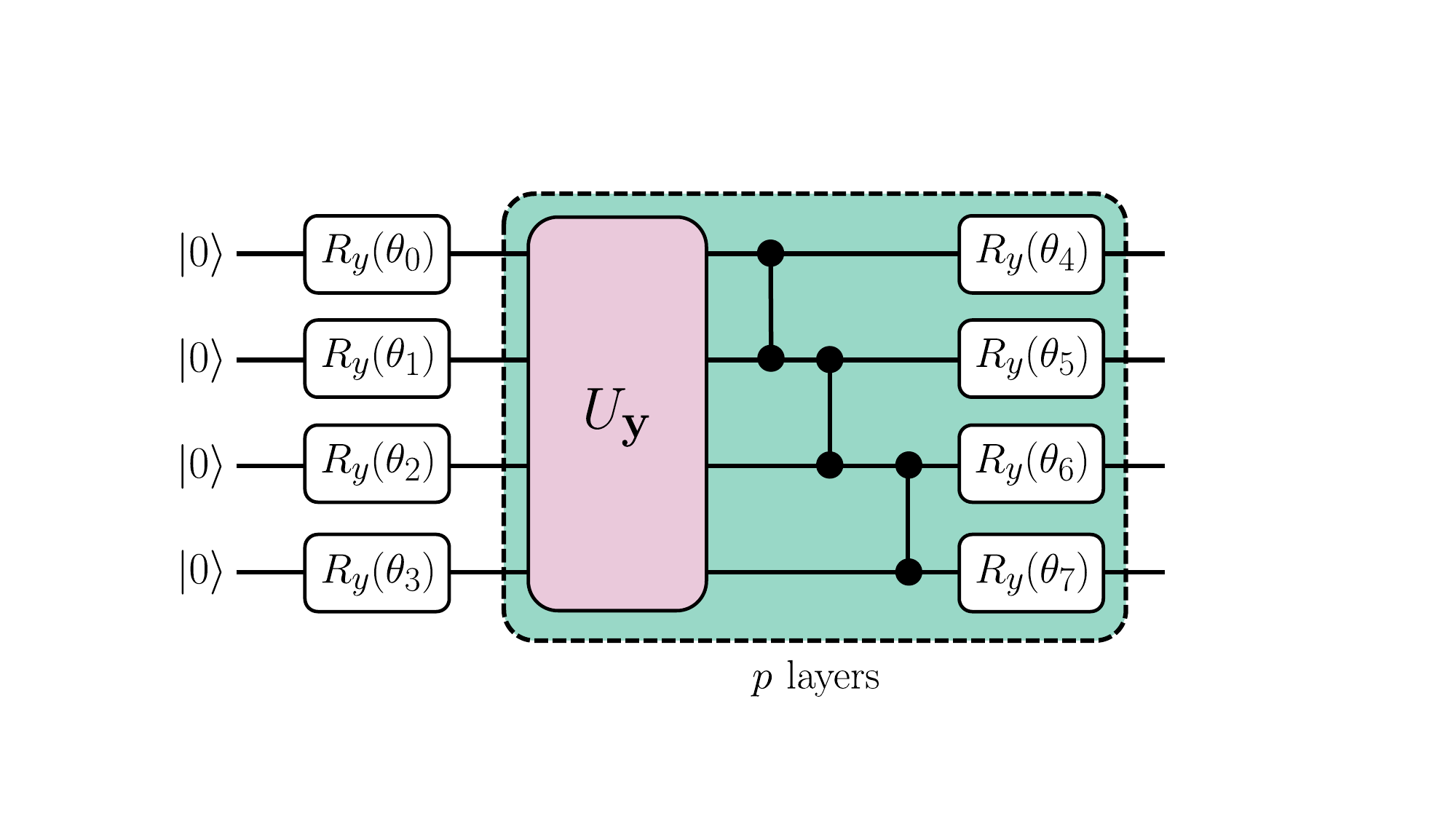}
    \caption{HEA with multiple data reuploadings (MUHEA) for a four-qubit example. The unitary $U_{\mathbf{y}}$ embeds the training labels $\mathbf{y}$ and is applied at the beginning of each layer. The ansatz consists of $n(p+1)$ trainable parameters $\boldsymbol{\theta}$, with $n$ qubits and $p$ layers}
    \label{fig:hea_many_ub}
\end{figure}

Additionally, we introduce a new ansatz that applies the unitary $U_{\mathbf{y}}$ in every layer before applying the CZ gates and rotational Y-gates, as depicted in Fig. \ref{fig:hea_many_ub}. This ansatz is inspired by the data reuploading techniques, shown in \cite{PerezSalinas2020datareuploading}. We call the ansatz which uses multiple unitary $U_{\mathbf{y}}$ operations MUHEA. 
In Section \ref{sec5_ansatz_comp} we carry out a comparison of the optimization behavior for each ansatz.
\subsection{Kernel} \label{sec4_kernels}
A crucial parameter for the approximation quality of GPs is the choice of the kernel function, as the expressivity of the GP model is fully determined by it. A kernel $k$ maps two inputs to a scalar output $k : \mathcal{X} \times \mathcal{X} \rightarrow \mathbb{K}$, where $\mathbb{K}$ is usually the space of real numbers $\mathbb{R}$ or complex numbers $\mathbb{C}$. For the sake of simplicity, we use the abbreviation $h=\| \mathbf{x} - \mathbf{x}'\|$ with $\mathbf{x}, \mathbf{x}' \in \mathcal{X}$ and $h\geq0$.

A widely used kernel function is the squared exponential or radial basis function (RBF) kernel
\begin{equation} \label{eq:kernel_rbf}
    k_{RBF}(h) = \sigma^2 \exp\left(- \frac{h^2}{2l^2}\right) \,,
\end{equation} 
with hyperparameters $\zeta = \{\sigma, l\}$, which can be interpreted as the amplitude $\sigma$ and the length scale $l$ with $\sigma,l \in \mathbb{R}^+$. These hyperparameters are typically optimized using maximum likelihood estimation. 

The choice of the kernel directly influences the VQLS convergence behavior. The kernel function is used to create the covariance matrix $\widehat{\mathbf{K}}$ which is the matrix of our linear system. This matrix is decomposed into a finite number of Pauli strings, as denoted in Eq. \eqref{eq:mat_decomp}. In the worst-case scenario, $L=N^2$ Pauli strings are needed. A higher number of Pauli strings also leads to a longer computation time. To reduce the number of Pauli strings, we use a kernel function that increases the number of zero components in the covariance matrix. 

We first introduce two types of kernels. 
For the space $\mathcal{X} \subseteq \mathbb{R}^d$ and constant $\nu \in \mathbb{R}^+$, the Mat\'ern kernel $k_{\nu} : \mathcal{X} \times \mathcal{X} \to \mathbb{R}$ is defined by  
\begin{equation} \label{eq:kernel_matern}
  k_{\nu}(h) = \sigma^2 \frac{1}{2^{\nu-1} \Gamma(\nu)} 
\left( \sqrt{2\nu} \frac{h}{l} \right)^{\nu} 
K_{\nu} \left( \sqrt{2\nu} \frac{h}{l} \right)  ,
\end{equation}
with the gamma function $\Gamma$ and the modified Bessel function of second kind $K_{\nu}$ \citep{matern2013spatial}. Similar to the RBF kernel, the Mat\'ern kernel depends on the hyperparameters $\zeta$, which includes the amplitude $\sigma$ and length scale $l$. The Mat\'ern kernel is a generalization of the RBF kernel, where the parameter $\nu$ controls the smoothness and as $\nu$ approaches infinity, the RBF kernel is obtained. A common choice for the parameter $\nu$ is $5/2$, which yields the special case
\begin{equation} \label{eq:kernel_matern_5/2}
k_{5/2}(h) = \sigma^2 \left(1 + \frac{\sqrt{5}h}{l} + \frac{5 h^2}{3 l^2}\right) \exp\left(-\frac{\sqrt{5}h}{l}\right) .
\end{equation}

We also introduce a kernel taper which was originally proposed by \cite{doi:10.1198/106186006X132178}. It is defined as 
\begin{equation} \label{eq:kernel_taper}
k_{\text{Taper}}(h) = \left(1 - \frac{h}{\theta}\right)_{+}^6 \left(1 + \frac{6 h}{\theta} + \frac{35 h^2}{3 \, \theta^2}\right)
\end{equation}
and denoted as the $\text{Wendland}_2$ taper. This kernel taper is compatible with the Mat\'ern kernel for smoothness values $\nu \leq 2.5$ and depends on the taper range $\theta$ which controls the extent of the tapering effect. We use the definition $x_+ = \max\{0, x\}$, which ensures a zero kernel value if the value $x$ is negative. Thus reducing the number of non-zero components in the covariance matrix, which also decreases the number of Pauli strings required to decompose the matrix $\mathbf{A}$. 

Following the approach of \cite{doi:10.1198/106186006X132178}, we combine these two kernels to obtain a new kernel 
\begin{equation} \label{eq:kernel_MT}
    k_{\text{MT}}(h) = k_{5/2}(h) \, k_{\text{Taper}}(h) ,
\end{equation}
which is the product of the Mat\'ern kernel and kernel taper, which we refer to as the MT kernel. We use the Mat\'ern kernel because the kernel taper is designed to be compatible only with kernel functions of the Mat\'ern family.
%%%%%%%%%%%%%%%%%%%%%%%%%%%%%%%%%%%%%%%%%%%%%%%%%%%%%%%%%%%%
%%% R E S U L T S %%%%%%%%%%%%%%%%%%%%%%%%%%%%%%%%%%%%%%%%%%
%%%%%%%%%%%%%%%%%%%%%%%%%%%%%%%%%%%%%%%%%%%%%%%%%%%%%%%%%%%%
\section{Results}\label{sec5} 
We evaluate the performance of the proposed VQLS-GP algorithm by performing regressions on a one-dimensional test function. We compare the accuracy of the given models for the RBF kernel and MT kernel, as presented in Section \ref{sec4_kernels}. Additionally, we investigate the advantages of using data reuploading techniques by comparing the three different ansätze presented in Section \ref{sec4_ansatzes}. 

The model accuracy is evaluated using the mean squared error (MSE) metric. We minimize the local cost function shown in Eq. \eqref{eq:local_cost_function} with a threshold $\text{tol} = 10^{-4}$ and a maximum number of $N_{\text{max}} = 1500$ iterations. We also use a maximum number of $2$ additional restarts for the VQLS optimization loop if the desired threshold is not reached. We do not perform hyperparameter optimization for the VQLS-GP models but rather compute the optimal hyperparameters with the GP models first and then reuse the optimized hyperparameters for the VQLS-GP models. The quantum circuits are implemented in PennyLane \citep{bergholm2022pennylaneautomaticdifferentiationhybrid}.
\begin{figure*}[ht]
    \centering
    \includegraphics[width=1.0\linewidth, trim=4.5cm 0.7cm 5.9cm 0.8cm, clip]{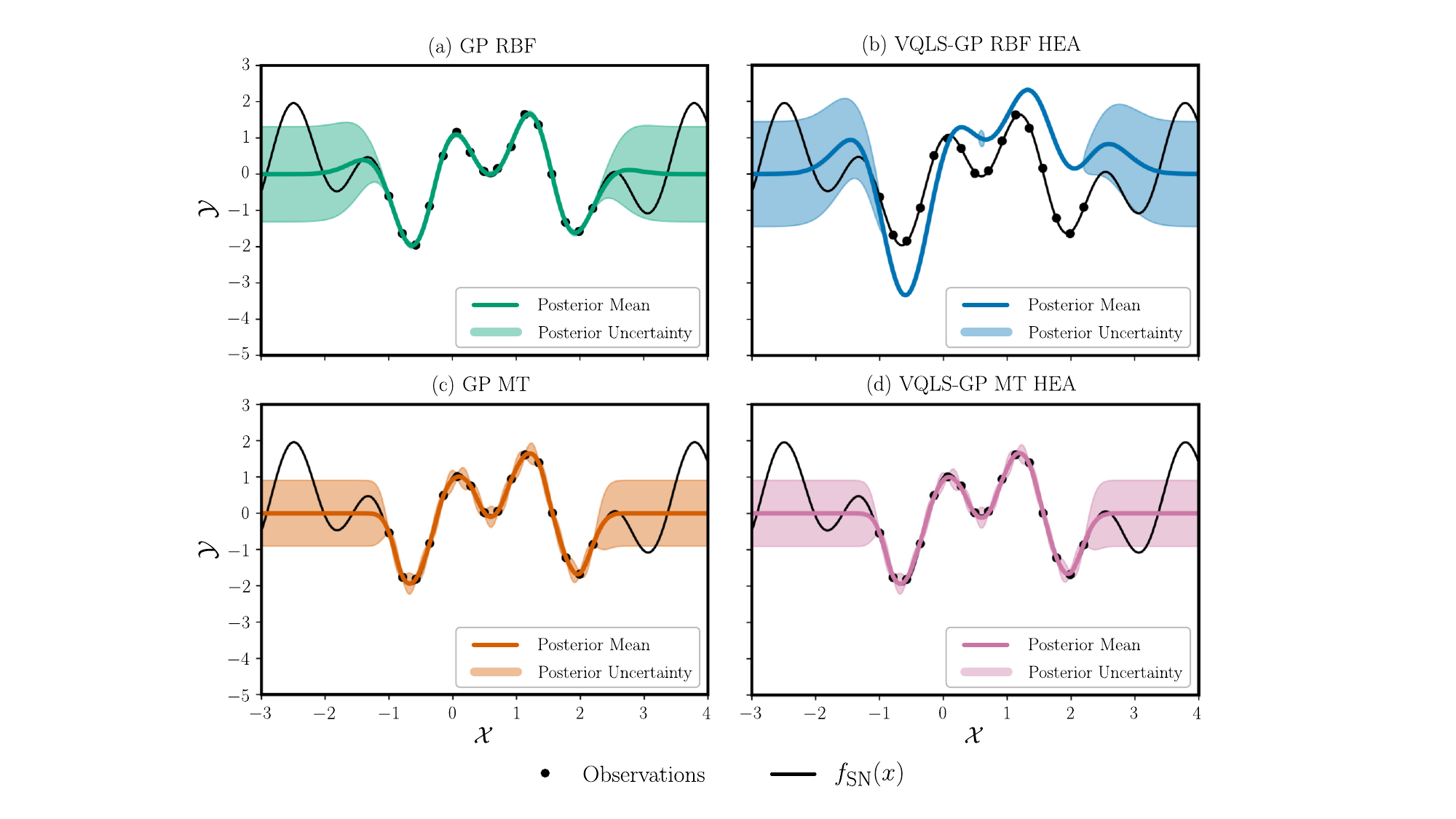}
    \begin{multicols}{2}
    \caption{GPR results for two GP and two VQLS-GP models applied to the Snelson test function (Eq. \eqref{eq:snelson}), comparing the regression quality of the RBF (Eq. \eqref{eq:kernel_rbf}) and MT (Eq. \eqref{eq:kernel_MT}) kernels. The plots display the training samples (black dots), the predictive mean (colored solid line), the predictive uncertainty (shaded region), and the latent function (solid black line). (a) shows the regression result of a classical GP model with an RBF kernel. (b) depicts the result of our VQLS-GP model with an RBF kernel. (c) shows a classic GPR with an MT kernel. (d) shows the result of our VQLS-GP model with an MT kernel. The hyperparameters are optimized within the classical GPRs and reused for the VQLS-GP models without any further modifications. Both VQLS-GP models use the HEA shown in Fig. \ref{fig:hea}}
    \label{fig:gpr_kernels}
    \end{multicols}
\end{figure*}
\begin{table*}
\caption{Kernel comparison for the GP and VQLS-GP models, averaged over $10$ evaluations. The number of training points is $N=16$. A total number of two additional restarts were allowed for each VQLS optimization. Only the best VQLS result was considered}\label{tab1}
\begin{tabular*}{\textwidth}{@{\extracolsep\fill}lccccc}
\toprule
Model & Kernel & Ansatz & Pauli strings & Iterations & MSE \\
\midrule
GP       & RBF &  -  &  -    & -              & $0.5780 \pm 0.0516$\\
GP       & MT  &  -  &  -    & -              & $0.5141 \pm 0.0010$\\
VQLS-GP  & RBF & HEA & $41$  & $24000 \pm 0$  & $6.1074 \pm 4.9725$\\
VQLS-GP  & MT  & HEA & $23$  & $21458 \pm 88$  & $0.5145 \pm 0.0010$\\
\botrule
\end{tabular*}
\end{table*}

For the regression problem, we use the following test function 
\begin{equation} \label{eq:snelson}
    f_{\text{SN}}(x) = \sin(2x) + \cos(5x)
\end{equation}
which is based on the Snelson dataset introduced in \cite{snelson2005sparse}. We use $N=16$ training points $\mathbf{X}$ sampled equidistantly from the interval $\left[-1.0, 2.2\right]$, and $M=1000$ test points $\mathbf{X}_*$ sampled equidistantly from the interval $\left[-3.0, 4.0\right]$. The model is trained using the Adam optimizer with a $\eta = 0.01$ learning rate \citep{kingma2014adam}. The function values $y_i$ for each $i={1,2,\dots,N}$ are evaluated from the objective function $f_\text{SN}$ with added Gaussian noise $\epsilon_i \sim \mathcal{N}(0, 0.1^2)$. 

We use $p=3$ layers for each HEA variation. For the VQLS-GP, we use $n+1$ qubits, where we need $n=\log_2{N} = 4$ qubits for the training data and one extra ancillary qubit to evaluate the Hadamard tests. Each model is evaluated a total of $10$ times.
\subsection{Kernel comparison} \label{sec5_kernel_comp}
We first analyze the choice of the kernel function and its impact on the regression results for both the GP and VQLS-GP models. In total, we compare two different kernels (RBF and MT) for two different models (GP and VQLS-GP). This results in a total of four model configurations for the kernel comparison. To focus solely on the influence of the kernel function, we use the standard HEA for the VQLS-GP models. For the MT kernel, we use a fixed taper range $\theta = 0.64$, which corresponds to $0.2$ times the length of the interval containing the training samples. 

The regression results with the smallest MSE values out of the $10$ repetitions for each model are shown in Fig. \ref{fig:gpr_kernels}. The regression result for the classical GP model with an RBF kernel is shown in Figure \ref{fig:gpr_kernels}(a), while the result of the VQLS-GP model with the same kernel is illustrated in Figure \ref{fig:gpr_kernels}(b). Regression results for both the classical GP and VQLS-GP using a MT kernel are presented in 
Figures \ref{fig:gpr_kernels}(c)-(d).

We maximized the LML as shown in Eq. \eqref{eq:lml} by minimizing its negative using an L-BFGS optimizer to find the optimal hyperparameters of each model \citep{bfgs}. The optimized hyperparameters for each regression result with the lowest MSE value are shown in Table \ref{tab2} in Appendix \ref{secA2}.

It is noticeable that the GP model with the RBF kernel has a slightly more accurate extrapolation behavior closely outside the region of the training data, as compared to the GP model with the MT kernel. Both the GP and VQLS-GP model using the MT kernel display a similar looking regression behavior. The VQLS-GP model with the RBF kernel has the most inaccurate regression fit, with the predictive mean deviating strongly from the true function. Computational errors, where the predictive variance yielded negative values due to non-convergence within the VQLS, resulted in the absence of uncertainties in Fig. \ref{fig:gpr_kernels}(b).

The uncertainties between the training points are higher for the models using the MT kernel, which may be explained by the locality of the kernel taper, as it incorporates information solely from a specified region near the regarded training point. Additionally, as the distance from the training data increases, the posterior mean curve becomes more horizontal and approaches a value of zero. This is caused by the kernel taper which disregards distance values above a certain threshold. 

The number of Pauli strings, total number of VQLS iterations, and  MSE values for the kernel comparison are summarized in Table \ref{tab1}. For the given regression problem, the VQLS-GP model using the RBF kernel requires a total amount of $41$ Pauli strings. Switching to the MT kernel reduces the amount of Pauli strings from $41$ to $23$, which is a decrease of approximately $44\%$. 

Since we need to solve $16$ linear systems, each with a maximum number of $1500$ iterations, the highest possible amount of total iterations is $24000$. The VQLS-GP model with the RBF kernel reaches this exact number, which means that the inner optimization loop of the VQLS did not convergence once. In contrast, the VQLS-GP model with the MT kernel only required an average of $21458$ iterations, indicating faster convergence. Exactly $12.5 \%$ of the VQLS optimization loops converged when the MT kernel was used, while not a single convergence was achieved using the RBF kernel. 

It is noticeable that the VQLS-GP using the RBF kernel has the highest MSE value with a mean of $6.1074$. The best MSE value is obtained with the GP using the MT kernel with a value of $0.5141$, followed by the VQLS-GP combined with the MT kernel with a MSE value of $0.5145$, which is almost identical to the best-performing model.

\begin{figure}[h]
    \centering
    \includegraphics[width=1.0\linewidth, trim=4.2cm 1.5cm 6.7cm 1.6cm, clip]{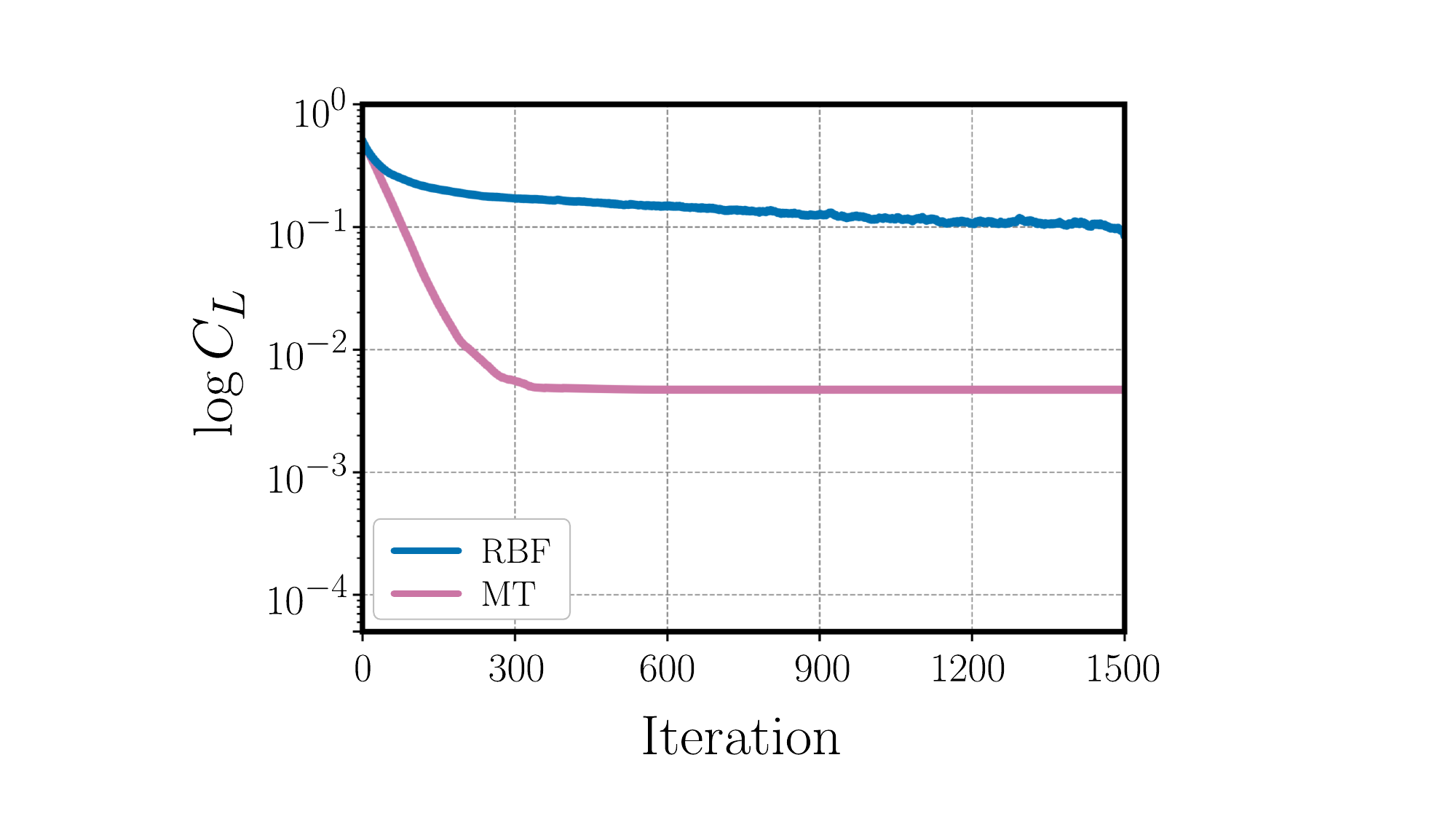}
    \caption{Averaged log-transformed loss curves for the VQLS-GP models of the kernel comparison in Section \ref{sec5_kernel_comp}. Each VQLS-GP run consists of $N=16$ VQLS optimization loops and each run was repeated $10$ times, resulting in a total of $160$ VQLS optimization loops for both the RBF and MT kernels. The corresponding regression results are depicted in Fig. \ref{fig:gpr_kernels} (b) and (d)}
    \label{fig:loss1}
\end{figure}
\begin{figure*}[h!]
    \centering
    \includegraphics[width=1.0\linewidth, trim=4.5cm 0.7cm 5.9cm 0.8cm, clip]{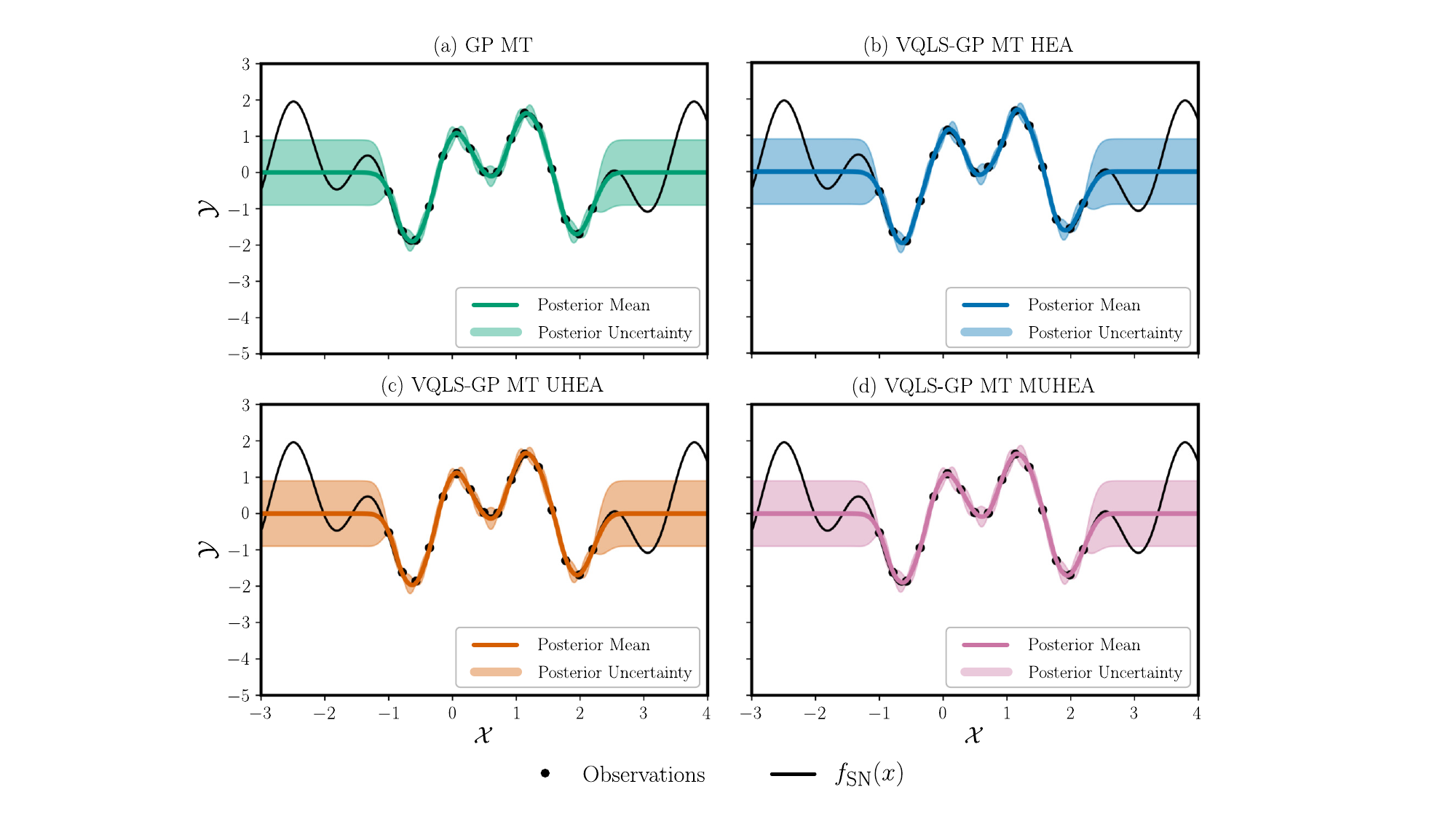}
    \begin{multicols}{2}
    \caption{GPR results for a classical GP and three VQLS-GP models applied to the Snelson test function (Eq. \eqref{eq:snelson}), comparing the regression quality of the HEA (Fig. \ref{fig:hea}), UHEA (Fig. \ref{fig:hea_ub}) and MUHEA (Fig. \ref{fig:hea_many_ub}). The plots display the training samples (black dots), the predictive mean (colored solid line), the predictive uncertainty (shaded region), and the latent function (solid black line). All four models use the MT kernel. (a) shows the regression result of a classical GP model. (b) depicts the result of our VQLS-GP model with a HEA. (c) shows a VQLS-GP model using the UHEA. (d) shows the result of our VQLS-GP model with the MUHEA. The hyperparameters are optimized within the classical GPRs and reused for the VQLS-GP models without any further modifications. All three VQLS-GP models use an MT kernel}
    \label{fig:gpr_ansatze}
    \end{multicols}
\end{figure*}
\subsection{Ansatz comparison} \label{sec5_ansatz_comp}
We now examine the impact of the different ansätze, as presented in Section \ref{sec4_ansatzes}, on the regression results while keeping the kernel fixed.
We use the MT kernel for both the classical GP and VQLS-GP models, which is motivated by the kernel comparison in Section \ref{sec5_kernel_comp}, where the MT kernel led to the best MSE values. To ensure consistency, we use the same test function as in the kernel comparison. 

The regression results for the ansatz comparison are depicted in Fig. \ref{fig:gpr_ansatze}. The posterior distribution of the GP model with the MT kernel is shown in Fig. \ref{fig:gpr_ansatze}(a). To analyze the impact of the ansatz choice, we evaluate the regression results of the VQLS-GP models using the HEA, UHEA and MUHEA ansätze, depicted in Fig. \ref{fig:gpr_ansatze}(b)-(d). These results demonstrate that all ansätze reveal nearly identical posterior distributions. The hyperparameters for each model are summarized in Table \ref{tab4} in Appendix \ref{secA2}.

We summarized the total number of iterations and MSE values for each model in Table \ref{tab3}. The model utilizing the UHEA is the least efficient in terms of iterations, as it failed to converge, resulting in a total of $24000$ iterations. In contrast, the model with the standard HEA required fewer iterations, with a mean value of $21522$ and also achieved a lower MSE value than the model using the UHEA. Among the three ansätze, the MUHEA required the fewest number of iterations with a mean of $10340$. Notably, using the MUHEA instead of the HEA reduces the number of iterations by approximately $52\%$. The model with the MUHEA also achieved the lowest MSE value among the VQLS-GP models. When comparing the classical GP model to the best performing VQLS-GP model, which uses the MUHEA, the MSE values are identical.

\begin{figure}[h]
    \centering
    \includegraphics[width=1.0\linewidth, trim=4.2cm 1.4cm 6.7cm 1.6cm, clip]{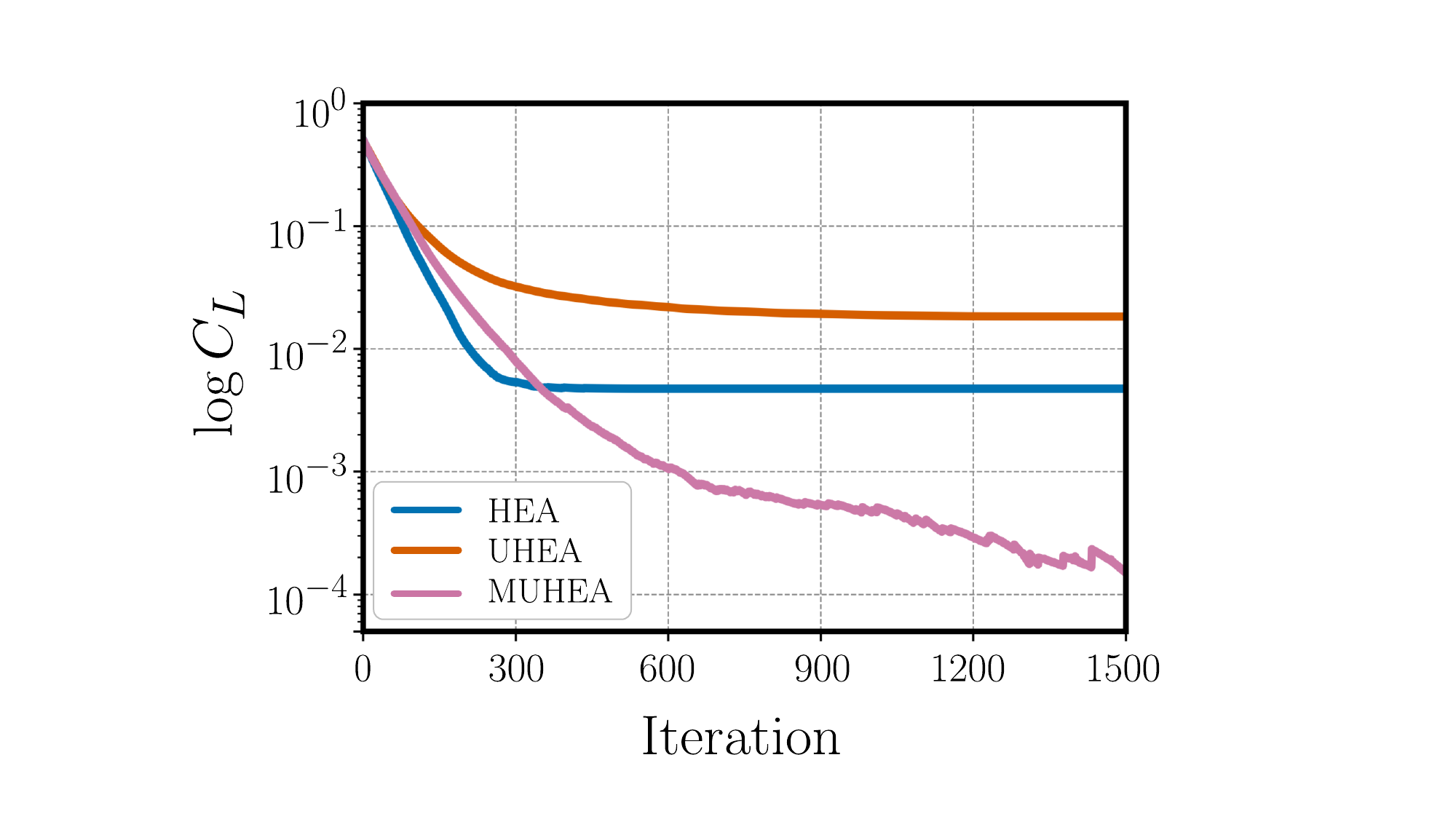}
    \caption{Averaged log-transformed loss curves for the VQLS-GP models of the ansatz comparison in Section \ref{sec5_ansatz_comp}. Each VQLS-GP run consists of $N=16$ VQLS optimization loops and each run was repeated $10$ times, resulting in a total of $160$ VQLS optimization loops for each of the HEA, UHEA and MUHEA ansätze. The corresponding regression results are depicted in Fig. \ref{fig:gpr_ansatze} (b)-(d)}
    \label{fig:loss2}
\end{figure}

The loss behavior for the VQLS-GP models in the ansatz comparison is depicted in Fig. \ref{fig:loss2}. Among the different ansätze, the UHEA resulted in the highest final average loss value, slightly above $10^{-2}$, and failed to converge in any of the simulations. In contrast, the model using the HEA performs comparably well with a final average loss value slightly below $10^{-2}$, and it converged in $12.5\%$ of all simulations. However, the best-performing ansatz is the MUHEA, which consistently reaches the convergence threshold of $10^{-4}$ in almost every optimization loop, achieving a convergence rate of $99.375 \%$.
\begin{table*}
\caption{Standard GP with MT kernel compared to VQLS-GP with MT kernel and three different HEA variations, averaged over $10$ evaluations. A total number of two additional restarts were allowed for each VQLS optimization. Only the best VQLS result was considered}\label{tab3}
\begin{tabular*}{\textwidth}{@{\extracolsep\fill}lccccc}
\toprule
Model & Kernel & Ansatz & Pauli strings & Iterations & MSE \\
\midrule
GP       & MT  &  -    &  -    & -                & $0.5145 \pm 0.0009$\\
VQLS-GP  & MT  & HEA   & $23$  & $21522 \pm 147$  & $0.5150 \pm 0.0009$\\
VQLS-GP  & MT  & UHEA  & $23$  & $24000 \pm 0$    & $0.5158 \pm 0.0014$\\
VQLS-GP  & MT  & MUHEA & $23$  & $10340 \pm 887$  & $0.5145 \pm 0.0009$\\
\botrule
\end{tabular*}
\end{table*}
%
%%%%%%%%%%%%%%%%%%%%%%%%%%%%%%%%%%%%%%%%%%%%%%%%%%%%%%%%%%%%
%%% D I S C U S S I O N %%%%%%%%%%%%%%%%%%%%%%%%%%%%%%%%%%%%
%%%%%%%%%%%%%%%%%%%%%%%%%%%%%%%%%%%%%%%%%%%%%%%%%%%%%%%%%%%%
\section{Discussion}\label{sec6} 
The scalability issues of GPR remain a critical challenge to its application in large-scale problems. The computational complexity of the covariance matrix inversion, which scales cubically with the  training data size, emphasizes the need for innovative methods to overcome this bottleneck. Variational quantum algorithms are promising candidates for addressing these scalability challenges by using both classical and quantum algorithms to accelerate computational tasks. 

This work presents a method to apply a hybrid quantum linear system solver to GPR problems. To be precise, we used the VQLS to compute a GP model's predictive mean and covariance. We presented two ways to do this. The first method is to compute the inverse of the covariance matrix by solving $N$ linear systems of equations with the VQLS, which can then compute the necessary terms for the posterior distribution. Additionally, we proposed a method that computes the matrix-vector and matrix-matrix products directly without computing the inverse of the covariance matrix by rewriting these terms into linear systems of equations.

The VQLS requires that the linear system matrix is decomposed into a set of unitaries, which can then be incorporated into the variational quantum circuits. A standard Pauli decomposition has an exponential complexity, making it intractable for higher sample sizes from a computational perspective. To overcome this problem, we combined a Mat\'ern kernel with a kernel taper to neglect less significant values in the covariance matrix, resulting in a lower number of unitaries and thus faster computation. For the Snelson test function example with $16$ training points, we reduced the number of Pauli strings by approximately $44\%$. 

We used the HEA and variations of this ansatz for the VQLS. To increase the expressivity of the ansatz, we added data reuploading techniques, which embed the right-hand side of the regarded linear system either once (UHEA) or in every layer (MUHEA). Compared to the standard HEA, the latter ansatz reduced the MSE value by $0.005$ and decreased the number of iterations by approximately $52\%$. In summary, the VQLS-GP model with the MT kernel and the MUHEA shows a comparable performance to the standard GP with the MT kernel.

We left out the hyperparameter optimization within the VQLS-GP algorithm to focus solely on the regression. Instead, we computed the optimized hyperparameters with the classical counterpart first and then reused them for the VQLS-GP. Furthermore, for the sake of simplicity, we only regarded one-dimensional objective functions. 

Future work could involve implementing the proposed algorithm using real quantum hardware to test the sensitivity and convergence behavior when dealing with noise. Additionally, the VQLS-GP could be combined with quantum kernels, potentially leading to an even better performance \citep{rapp2024quantum}. More complex examples with higher dimensions could be tested. Furthermore, examples with more training data and the occurrence of barren plateaus could be analyzed. 

%%%%%%%%%%%%%%%%%%%%%%%%%%%%%%%%%%%%%%%%%%%%%%%%%%%%%%%%%%%%
%%% A P P E N D I X %%%%%%%%%%%%%%%%%%%%%%%%%%%%%%%%%%%%%%%%
%%%%%%%%%%%%%%%%%%%%%%%%%%%%%%%%%%%%%%%%%%%%%%%%%%%%%%%%%%%%
\begin{appendices}
\counterwithout{figure}{section} % Remove section dependency from figure numbering
\setcounter{figure}{7}% Set the figure counter to the last figure number in the main text
\counterwithout{table}{section} % Remove section dependency from figure numbering
\setcounter{table}{2} % Set the figure counter to the last figure number in the main text

\section{Hadamard test}\label{secA1}
\begin{figure}[htbp] 
    \centering
    \includegraphics[width=\linewidth, trim=4cm 5cm 6cm 5cm, clip]{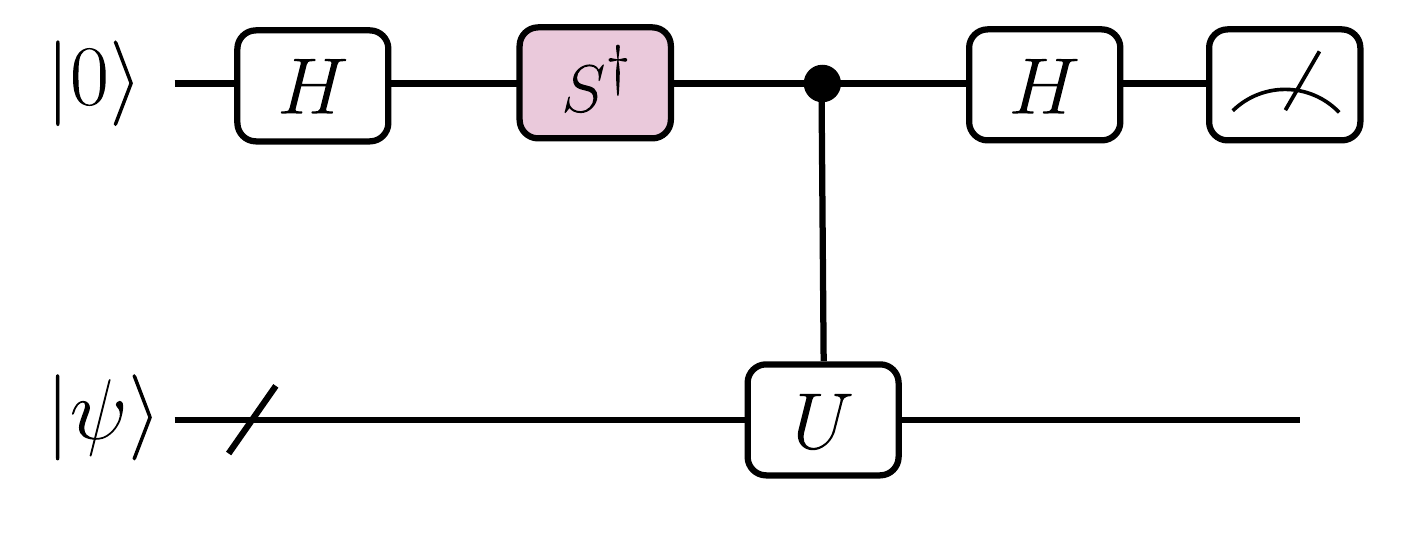}
    \caption{Quantum circuit structure of the Hadamard test. The $S^\dagger$ gate is only used when the imaginary part of the expectation value is computed}
    \label{fig:hadamard}
\end{figure}

The Hadamard test is a fundamental quantum circuit with a very simple structure. Let $\ket{\psi}$ be a quantum state, and let $U$ be an arbitrary unitary operation applied to this state. The Hadamard test computes the expectation value $\expval{U}{\psi}$. The quantum circuit is shown in Fig. \ref{fig:hadamard}. 
% \begin{figure}[htbp] 
%     \centering
%     \includegraphics[width=\linewidth, trim=4cm 5cm 6cm 5cm, clip]{Fig8.pdf}
%     \caption{Quantum circuit structure of the Hadamard test. The $S^\dagger$ gate is only used when the imaginary part of the expectation value is computed}
%     \label{fig:hadamard}
% \end{figure}
\noindent
An ancillary qubit must first be prepared to the $\ket{0}$ state. Additionally, the system qubits must be initialized in the state $\ket{\psi}$, resulting in the initial state
\begin{equation}
   \ket{0} \otimes \ket{\psi} .
\end{equation}
This state is then modified by applying a Hadamard gate to the ancillary qubit, resulting in a superposition state 
\begin{equation}
    \frac{1}{\sqrt{2}}(\ket{0} + \ket{1}) \otimes \ket{\psi}.
\end{equation}
In the quantum circuit shown in Fig. \ref{fig:hadamard}, a controlled unitary $U$ is applied, meaning that $U$ is applied to the state $\ket{\psi}$ if the ancillary qubit is in the state $\ket{1}$. The global state takes the form 
\begin{equation}
    \frac{1}{\sqrt{2}}(\ket{0} \otimes \ket{\psi} + \ket{1} \otimes U\ket{\psi}).
\end{equation}
The next step is to apply a Hadamard gate to the ancillary qubit again. After further simplifications, the resulting state takes the following form
\begin{equation}
    \ket{\Phi} = \frac{1}{2} \left( \ket{0} \otimes (\ket{\psi} + U\ket{\psi}) + \ket{1} \otimes (\ket{\psi} - U\ket{\psi}) \right).
\end{equation}
The final step involves measuring the ancillary qubit, which enables the computation of the desired expectation value. 

To measure the probability of the ancillary qubit being in the zero state, we apply the operator $P = \ketbra{0}{0} \otimes \mathbb{I}$ to the quantum state $\ket{\Phi}$. The probability of obtaining the $\ket{0}$ state is given by computing $P(0) = \|P\ket{\Phi}\|^2$. The operator $\ketbra{0}{0}$ ensures that the $\ket{0}$ state remains and that the $\ket{1}$ state is cancelled out. 

Using $P = \ketbra{0}{0} \otimes \mathbb{I}$, the probability of measuring the zero state is
\begin{equation}
    P(0) = \left\| P \ket{\Phi} \right\|^2 = \frac{1}{2} \left( 1 + \real \expval{U}{\psi} \right) .
\end{equation}
To obtain the real part of the expectation value, we can reformulate the above equation as follows
\begin{equation}
    \real \expval{U}{\psi} = 2 P(0) - 1 .
\end{equation}
For the imaginary part of the expectation value, a simple adjoint phase gate $S^\dagger$ is applied after the first Hadamard gate. This modification yields the imaginary part
\begin{equation}
    \imaginary \expval{U}{\psi} = 2 P(0) - 1 .
\end{equation}
The full expectation value can be obtained by combining both the real and imaginary parts, resulting in
\begin{equation}
  \expval{U}{\psi} = \real \expval{U}{\psi} + i \,  \imaginary \expval{U}{\psi} .
\end{equation}
\section{Hyperparameters}\label{secA2}
The hyperparameters of the models with the regression results that are shown in Figs. \ref{fig:gpr_kernels} and \ref{fig:gpr_ansatze} are presented in Tables \ref{tab2} and \ref{tab4} respectively. 
\begin{table}[htbp]
  \centering
  \normalsize
  \caption{Hyperparameters for the kernel comparison, presented in Section \ref{sec5_kernel_comp}. Corresponding regression plots are shown in Fig. \ref{fig:gpr_kernels}}\label{tab2}
  \begin{tabular}{p{0.62\linewidth} c c}
    \toprule
    Model Configuration & $\sigma$ & $l$ \\
    \midrule
    GP RBF & $1.3133$ & $0.3964$ \\
    GP MT & $0.9083$ & $1.2417$ \\
    VQLS-GP RBF HEA & $1.4489$ & $0.4322$ \\
    VQLS-GP MT HEA & $0.9083$ & $1.2417$ \\
    \botrule
  \end{tabular}
\end{table}
\begin{table}[htbp]
  \centering
  \normalsize
  \caption{Hyperparameters for the ansatz comparison, presented in Section \ref{sec5_ansatz_comp}. Corresponding regression plots are shown in Fig. \ref{fig:gpr_ansatze}}\label{tab4}
  \begin{tabular}{p{0.62\linewidth} c c}
    \toprule
    Model Configuration & $\sigma$ & $l$ \\
    \midrule
    GP MT & $0.9015$ & $2.5394$ \\
    VQLS-GP MT HEA& $0.9014$ & $2.6829$ \\
    VQLS-GP MT UHEA & $0.9015$ & $2.5394$ \\
    VQLS-GP MT MUHEA & $0.9015$ & $2.5394$ \\
    \botrule
  \end{tabular}
\end{table}
\end{appendices}
%
%%%%%%%%%%%%%%%%%%%%%%%%%%%%%%%%%%%%%%%%%%%%%%%%%%%%%%%%%%%%
%%% D E C L A R A T I O N S %%%%%%%%%%%%%%%%%%%%%%%%%%%%%%%%
%%%%%%%%%%%%%%%%%%%%%%%%%%%%%%%%%%%%%%%%%%%%%%%%%%%%%%%%%%%%
\backmatter

\vspace{0.2cm}
\noindent
{\small{\textbf{Acknowledgements} We would like to express our sincere gratitude to Daniel Grealy and Krzysztof Bieniasz for their valuable contributions.
}}
\newline
\newline
\noindent
% {\small{\textbf{Author contribution} Kerem Bükrü: Conceptualization, Formal analysis, Methodology, Software, Visualization, Writing - original draft. Steffen Leger: Data curation, Methodology, Software, Writing - review \& editing.
% M. Lautaro Hickmann: Methodology, Writing - review \& editing. Hans-Martin Rieser: Funding acquisition, Supervision. Ralf Sturm: Funding acquisition, Supervision. Tjark Siefkes: Supervision.}} 
{\small{\textbf{Author contribution} K.B. conceptualized the study, performed formal analysis, developed methodology, implemented software, created visualizations, and wrote the original draft. S.L. curated data, developed methodology, implemented software, and reviewed and edited the manuscript. M.L.H. developed methodology and reviewed and edited the manuscript. H.-M.R. and R.S. acquired funding and supervised the project. T.S. supervised the project.}} 

\section*{Declarations}
{\small{\textbf{Conflict of interest} The authors declare no competing interests.}
\newline
\newline
\noindent
\small{\textbf{Funding} Kerem B\"ukr\"u and M. Lautaro Hickmann received funding from the DLR Quantum Fellowship Program. }}

\begin{comment}
\bmhead{Open Access} This article is licensed under a Creative Commons Attribution 4.0 International License, which permits use, sharing, adaptation, distribution and reproduction in any medium or format, as long as you give appropriate credit to the original author(s) and the source, provide a link to the Creative Commons licence, and indicate if changes were made. The images or other third party material in this article are included in the article’s Creative Commons licence, unless indicated otherwise in a credit line to the material. If material is not included in the article’s Creative Commons licence and your intended use is not permitted by statutory regulation or exceeds the permitted use, you will need to obtain permission directly from the copyright holder. To view a copy of this licence, visit http:// creativecommonshorg/licenses/by/4.0/.
\end{comment}

%%%%%%%%%%%%%%%%%%%%%%%%%%%%%%%%%%%%%%%%%%%%%%%%%%%%%%%%%%%%
%%% B I B L I O G R A P H Y %%%%%%%%%%%%%%%%%%%%%%%%%%%%%%%%
%%%%%%%%%%%%%%%%%%%%%%%%%%%%%%%%%%%%%%%%%%%%%%%%%%%%%%%%%%%%
% \FloatBarrier
\bibliography{sn-bibliography}

\end{document}